\documentclass[preprint]{aastex631}
\usepackage{amsmath}

\shorttitle{Polarization Angle Geodesics}
\shortauthors{McKINNON}
\begin{document}

\title{Polarization Angle Geodesics in PSRs B1133+16 and B2016+28}
\author{M. M. McKinnon}
\affiliation{National Radio Astronomy Observatory, Socorro, NM \ 87801 \ \ USA}
\correspondingauthor{M. M. McKinnon}
\email{mmckinno@nrao.edu}


\begin{abstract}
Recent models of pulsar polarization predict that the position and ellipticity 
angles of the polarization vector can trace portions of a small or great circle on 
the Poincar\'e sphere. A great circle can arise from a transition in dominance of 
orthogonal polarization modes, where the relative intensity of the modes changes 
with pulse longitude. A small circle may be caused by a rotation of the vector, 
where the phase difference between the modes changes with pulse longitude or 
wavelength. Observations of PSRs B1133+16 and B2016+28 are reanalyzed to search for 
these polarization features within their pulse profiles. The polarization angles 
observed in part of PSR B1133+16 are shown to follow a great circle on the Poincar\'e 
sphere. The angles observed across the pulse of PSR B2016+28 follow an arc that 
resembles a portion of a great circle that has been altered by the pulsar's rotation. 
The observations are interpreted within the context of three different polarization 
models. All three models produce similar results for both pulsars and indicate that 
the observed geodesics are caused by mode transitions. The arc observed in PSR 
B2016+28 can also be interpreted as a vector rotation, provided the modes are 
elliptically polarized. The observations and accompanying analysis show that mode 
transitions are not restricted to the equatorial plane of the Poincar\'e sphere and 
that arcs and partial circles may be more common than previously recognized. 
\end{abstract}


\section{Introduction}

The radio emission of pulsars frequently exhibits abrupt discontinuities in position angle
(PA, $\psi$) of $\Delta\psi\sim 90^\circ$. In the instances where the ellipticity angle 
(EA, $\chi$) of the polarization vector has also been reported, the PA discontinuities are 
often accompanied by large excursions in the EA, as observed in PSR B0031-07 (C. D. Ilie 
et al. 2020), PSR B0329+54 (R. T. Edwards \& B. W. Stappers 2004, hereafter ES04), PSR 
B0809+74 (R. T. Edwards 2004), and PSR J1157-6224 (L. S. Oswald et al. 2023, hereafter OKJ). 
Historically, the PA discontinuities have been interpreted as transitions between modes of 
orthogonal polarization (OPMs), where the relative intensity of the modes changes with 
pulse longitude (e.g., R. N. Manchester et al. 1975; J. M. Cordes et al. 1978; D. R. 
Stinebring et al. 1984, hereafter S84). However, J. Dyks et al. (2021, hereafter DWI) 
suggested that large changes in both the PA and EA may arise from a rotation of the 
polarization vector, where the phase difference between the modes changes with wavelength 
or longitude, as might be expected from generalized Faraday rotation or Faraday conversion 
in the pulsar's magnetosphere or wind (W. J. Cocke \& A. G. Pacholczyk 1976; D. B. Melrose 
1979; M. Kennett \& D. Melrose 1998; ES04; M. Lyutikov 2022; M. E. Lower et al. 2024). 
They developed a polarization model that highlights the observational differences between 
a mode transition and a vector rotation. Their model shows that the PA-EA pairs of a mode 
transition trace a portion of a great circle (GC) on the Poincar\'e sphere, whereas the 
pairs of a vector rotation generally trace a small circle (SC) on the sphere. Polarization 
arcs and partial circles have been observed in fast radio bursts (A. Bera et al. 2025), 
long-period radio transients (D. Dobie et al. 2024, J. Pritchard et al. 2026), a magnetar 
(M. E. Lower et al. 2024), and some pulsars (e.g., OKJ; S. Cao et al. 2025). 

The DWI polarization model assumes the radio emission is composed of two coherent, 
linearly polarized OPMs and consequently requires the emission to be completely polarized, 
which generally is not observed. OKJ addressed this issue by expanding the model to include 
partially coherent OPMs (PCOH). Following suggestions by S84 and ES04, M. M. McKinnon (2024, 
hereafter M24) developed two additional models of polarization mode transitions. One model 
assumes the polarization modes are incoherent and nonorthogonal (NPMs). The other model 
assumes the OPMs are incoherent and accompanied by an elliptically polarized emission 
component (EPC). He showed that all four models of mode transitions predict that the PA-EA 
pairs trace a GC geodesic that connects the polarization vectors on the surface of the 
Poincar\'e sphere. 
 
OKJ introduced a diagram that facilitates the interpretation of the observed values of 
the total polarization fraction, $p$, and the absolute value of the polarization vector's 
latitude, $\theta=|\lambda|=|2\chi|$, within the context of their PCOH model. The model 
is characterized by three physical parameters (see Section~\ref{sec:pcoh}). By holding 
two of the parameters constant and allowing the third to vary, one can use the model and 
the diagram to illustrate how $p$ and $\theta$ should change with the varying parameter 
(see their Figure 1). The observed variations in $p$ and $\theta$ can then be compared 
with the theoretical $p$-$\theta$ tracks in the diagram to investigate their physical 
origin. In doing so, OKJ found that the $p$-$\theta$ variations observed in parts of PSRs 
J0820-1350 and J1157-6224 were consistent with vector rotations, while the variations 
in a part of PSR J0134-2937 were consistent with a mode transition. In general, however, 
the observational results and their subsequent interpretation can be more complex, 
because in principle, all three model parameters can vary. And since there are two 
observable quantities, $p$ and $\theta$, and three unknown parameters, OKJ's analysis 
can be augmented to investigate cases where two model parameters are allowed to vary 
while one remains constant.

The present paper has two objectives. The first objective is to identify additional 
examples of polarization arcs and partial circles by reanalyzing existing polarization 
observations of two pulsars that are known to have large PA discontinuities in their 
pulse profiles: PSRs B1133+16 and B2016+28. Evidence for polarization arcs and circles 
is found in both pulsars. The second objective is to determine if these polarization 
features are due to mode transitions or vector rotations and to ascertain if one 
polarization model is more representative of the observations than the others. 

The paper is organized as follows. The polarization observations of PSRs B1133+16 and 
B2016+28 made by S84 are reanalyzed in Section~\ref{sec:obs}, primarily to include the 
EA. The PCOH, EPC, and NPM models of pulsar polarization are briefly summarized in 
Section~\ref{sec:models} to describe how the observed EA and polarization fraction are 
affected by the parameters of each model. The observations are then interpreted within 
the context of these models in Sections~\ref{sec:1133} and~\ref{sec:2016}. The analysis 
is based on the approach adopted by OKJ, where the model parameters are estimated from 
the measured values of $p$ and $\chi$. The primary difference between the analysis and 
that of OKJ is two parameters are allowed to vary in this analysis, whereas only one 
parameter varies in OKJ. The results of the analysis are discussed in 
Section~\ref{sec:discuss}, and conclusions are summarized in Section~\ref{sec:conclude}. 
Appendix~\ref{sec:circles} illustrates the vector geometries for mode transitions and 
vector rotations and lists the equations for an SC and a GC. Appendix~\ref{sec:parms} 
lists the equations for the polarization model parameters as functions of the observed 
quantities $p$ and $\chi$.


\section{Observational Results}
\label{sec:obs}

\begin{figure}
\plotone{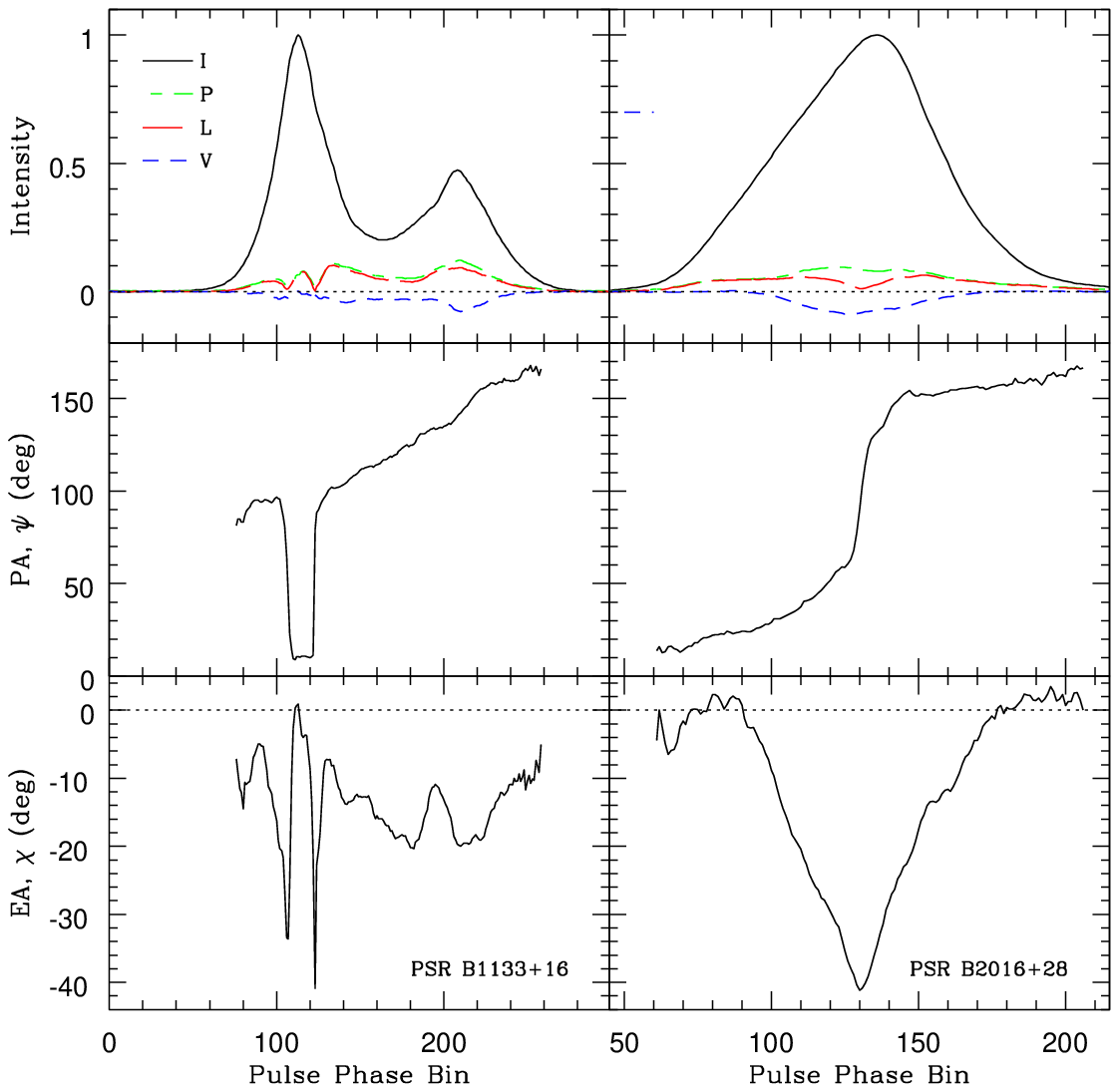}
\caption{Pulse profiles of PSR B1133+16 (left column) and PSR B2016+28 (right column) at 
1404 MHz. The top row of the figure shows the total intensity (I, solid black line), total 
polarization (P, dashed green line), linear polarization (L, dashed red line), and circular
polarization (V, dashed blue line) for each pulsar. The variations in the PAs and EAs 
across the pulse of each pulsar are, respectively, shown in the middle and bottom rows of 
the figure. The profiles have been reproduced from Figures 16 and 30 of S84.} 
\label{fig:profiles}
\end{figure}

\subsection{Pulse Profiles}

S84 conducted single-pulse polarization observations of 11 bright, low dispersion measure 
pulsars with the Arecibo radio telescope with the primary goals of investigating the 
frequency dependence and longitudinal extent of OPMs within their pulse profiles. Their 
observations of PSRs B1133+16 and B2016+28 were made at a frequency of 1404 MHz centered 
on a bandwidth of 10 MHz. The data sampling interval for both pulsars was 0.2 ms. S84 
went to great lengths to properly calibrate their observations and to understand and 
minimize sources of instrumental error. Therefore, the effects of calibration and other 
instrumental errors upon the observational results are deemed to be minor. The details 
of the data collection, calibration, and analysis are discussed in the original paper.

S84 did not report measurements of the pulsars' EAs. Their pulse profiles of PSRs 
B1133+16 and B2016+28 are reproduced in Figure~\ref{fig:profiles} with the EAs included. 
The left column of the figure shows the profile of PSR B1133+16, and the right column 
shows the profile of PSR B2016+28. The top row of the figure shows the total intensity 
($I$), total polarization ($P$), linear polarization ($L$), and circular polarization 
($V$) for each pulsar's profile. The linear, $L=(Q^2+U^2)^{1/2}$, and total polarization, 
$P=(Q^2+U^2+V^2)^{1/2}$, were compensated for contributions from the instrumental noise
using the general purpose polarization estimators described in M. M. McKinnon (2025a). 
The data in these panels have been normalized to the peak total intensity of each profile. 
The middle and bottom rows of the figure show how the PA and EA, respectively, vary 
across the profiles. The PA and EA were calculated with
\begin{equation}
\psi = \frac{1}{2}\arctan{\left(\frac{\langle U\rangle}{\langle Q\rangle}\right)},
\label{eqn:PAdefn}
\end{equation}
\begin{equation}
\chi = \frac{1}{2}\arctan{\left(\frac{\langle V\rangle}{\langle L\rangle}\right)},
\label{eqn:chi}
\end{equation}
where the angular brackets denote averages over multiple pulses at a given pulse longitude. 
The PA values are not absolute.

The profile of PSR B1133+16 consists of two components. The polarization across the pulse is
generally low. Two notches in the total and linear polarization appear in the leading component.
The notches are accompanied by $\sim 90^\circ$ discontinuities in the PA and abrupt excursions 
in the EA. An EA excursion occurs at each notch, because the magnitude of the circular 
polarization exceeds the linear polarization at those locations (see Equation~\ref{eqn:chi}). 
After the PA discontinuities, the PA increases roughly linearly across the pulse. The EA 
gradually decreases, then peaks at $\chi=-10.9^\circ$ near the trailing pulse component, 
before gradually increasing over the remainder of the pulse.

The profile of PSR B2016+28 consists of a single broad component. The polarization across 
the pulse is low. A notch in the linear polarization precedes the pulse's peak in total
intensity. The magnitude of the circular polarization is near maximum at the notch, and the 
total polarization remains approximately constant over this region of the pulse. A large 
discontinuity in the PA occurs at the notch in linear polarization. The PA elsewhere varies 
roughly linearly across the pulse. The EA consists primarily of a broad, V-shaped feature. The 
vertex of the V coincides with the linear polarization notch and the PA discontinuity. The EAs 
in the sides of the V vary linearly with pulse phase bin (ppb) from ppb 94 down to ppb 130 and 
back up to ppb 153.


\subsection{Trajectories of the Polarization Angles}

\begin{figure}
\plotone{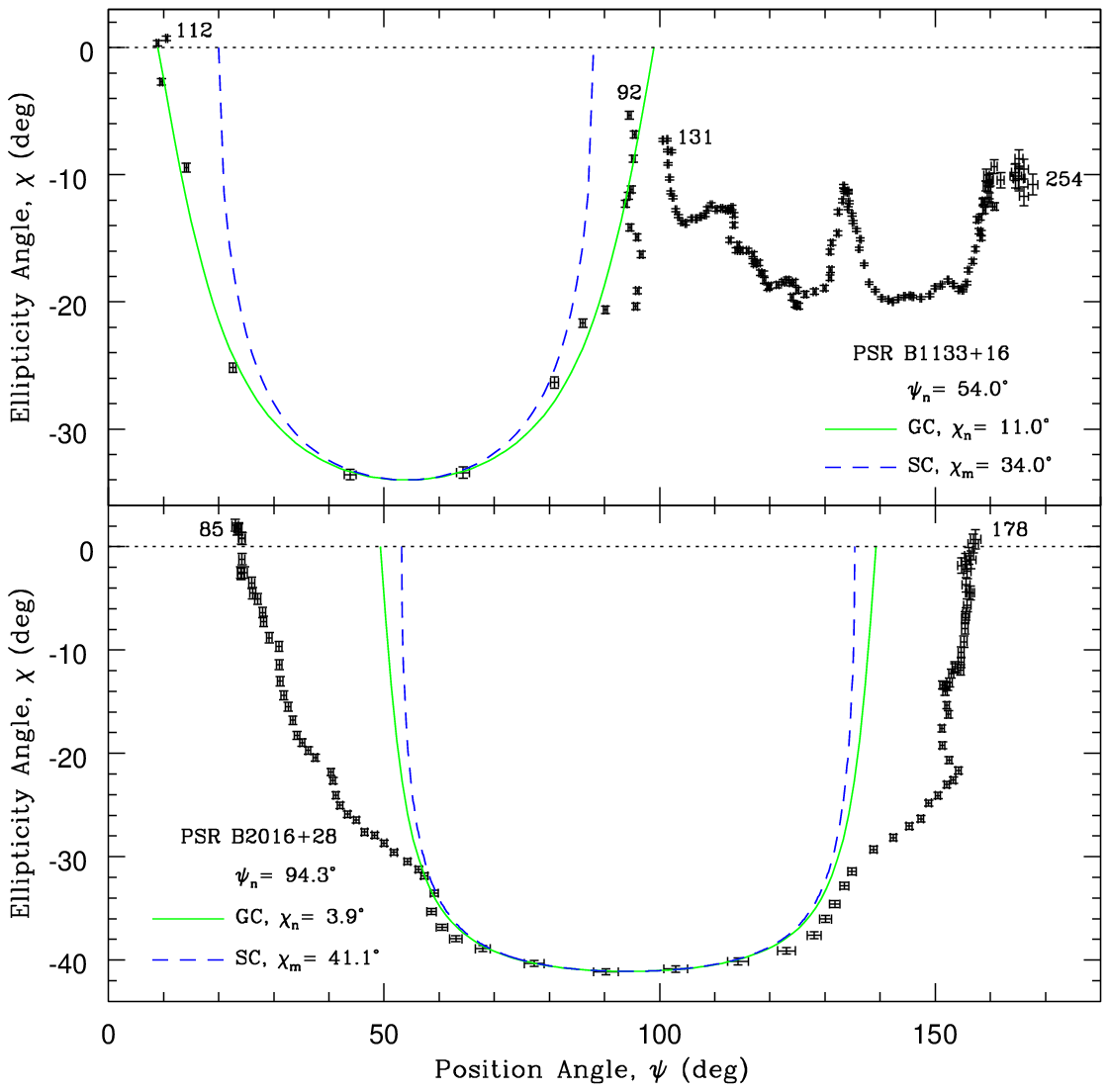}
\caption{Dependence of pulsar EAs upon their PAs. The top panel shows the PA-EA pairs observed
in PSR B1133+16. The bottom panel shows the pairs observed in PSR B2016+28. Small circles (SC, 
dashed blue lines) and great circles (GC, solid green lines) are overlaid on portions of the 
data in both panels. The parameters that characterize the circles are annotated in each panel. 
The isolated numbers in the panels denote the pulse phase bins associated with their adjacent 
data points.}
\label{fig:EAPA}
\end{figure}

Figure~\ref{fig:EAPA} shows how the pulsars' EAs vary as functions of their PAs in portions 
of their pulse profiles. The formal errors assigned to the polarization angle measurements 
are $68\%$ confidence limits calculated from
\begin{equation}
\sigma_{\psi,\chi} = \frac{1}{2s} = \frac{28.65^\circ}{s},
\label{eqn:angerr}
\end{equation}
where $s$ is the signal-to-noise ratio (SNR) in linear polarization for the PA confidence limit,  
$\sigma_\psi$, and is the SNR in total polarization for the EA confidence limit, $\sigma_\chi$ 
(M. M. McKinnon 2025b). Equation~\ref{eqn:angerr} is generally an accurate estimate of the 
confidence limits when the polarization SNR at each ppb is large, as is the case for both 
pulsars. Since the SNR in total polarization is greater than or equal to the SNR in linear 
polarization, the EA confidence limit is always less than or equal to the PA confidence 
limit, $\sigma_\chi\le\sigma_\psi$. 

The top panel of Figure~\ref{fig:EAPA} shows the PA-EA pairs of PSR B1133+16 over ranges in 
ppb of 92-112 and 131-254. The ppb numbers are annotated in the panel. The PA-EA pairs in 
the ppb range of 92-112 occur over the first PA discontinuity in the profile, and the pairs 
in the range of 131-254 occur after the second PA discontinuity. The solid green line and 
dashed blue line overlaid on the PA-EA pairs in the ppb range of 92-112 represent portions 
of a GC and an SC, respectively. The geometrical parameters of the GC and the SC are 
annotated in the panel (see Figure~\ref{fig:gcsc} and Equations~\ref{eqn:gc} and~\ref{eqn:sc}). 
The trajectory of the PA-EA pairs is qualitatively consistent with a GC but not a SC. From the 
middle-left panel of Figure~\ref{fig:profiles}, the PAs immediately preceding and following the 
PA discontinuity are approximately constant, suggesting they are not significantly affected 
by the rotation of the star (e.g., via the rotating vector model, RVM, of V. Radhakrishan \& 
D. J. Cooke 1969) in this region of the pulse and consequently preserve the near-pristine GC 
trajectory followed by the PA-EA pairs. In the ppb range of 131-254, the EA initially decreases 
with increasing PA and then increases toward a distinct peak of $\chi=-10.9^\circ$ at a PA of 
$\psi=133.4^\circ$. The EA peak is not accompanied by a PA discontinuity, and its shape is not 
consistent with a GC or SC. The PA-EA pairs covering the second PA discontinuity in the profile 
are not shown in the panel for clarity. They also appear to follow a GC; however, the time 
resolution in this region of the pulse is insufficient to adequately constrain this GC's 
geometry.

The bottom panel of Figure~\ref{fig:EAPA} shows how the PA-EA pairs of PSR B2016+28 vary over 
most of its pulse. The data points span a ppb range of 85-178. The pairs form a large arc with 
a nadir in EA of $\chi=-41.1^\circ$ occurring at a PA of $\psi=94.3^\circ$. The solid green line 
and dashed blue line shown in the panel represent a GC and an SC aligned with the center of the 
arc. The circles are consistent with the data points at the arc center, but are inconsistent 
with the data elsewhere. The discrepancy between the circles and the PA-EA pairs may be due to 
the PA variations caused by the pulsar's rotation, because the average PA on the pulse edges
is not constant, but instead varies roughly linearly with ppb (Figure~\ref{fig:profiles}). The 
PA tracks of the individual polarization modes also vary linearly across the entire pulse (see 
Figure 31 of S84 and Figure 1 of M. M. McKinnon 2003, hereafter M03). The overall effect of 
the pulsar's rotation is to extend the range of the GC's PA via the RVM (e.g., see Section 3.3 
of M24). 

The observations of PSRs B1133+16 and B2016+28 show that PA-EA pairs in parts of their 
profiles trace portions of GC-like features. What remains to be determined is whether the 
PA-EA variations are caused by mode transitions or vector rotations and which 
polarization model best describes the observations. 


\section{Summary of Polarization Models}
\label{sec:models}

Each of the PCOH, EPC, and NPM models of pulsar polarization is characterized by three 
physical parameters. Equations for the emission's polarization fraction and EA can be 
written as functions of these parameters (M24), as summarized in what follows. The 
parameters generally differ between models. However, one parameter, the ratio of the 
mode mean intensities, is common to all models. If the two OPMs are designated as A 
and B and their intensities at a given pulse longitude are defined by the random 
variables $X_A$ and $X_B$, the ratio of the mode mean intensities is 
$M=\langle X_A\rangle/\langle X_B\rangle$ (M. M. McKinnon 2022; M24). The equations 
for $p$ and $\chi$ can be simplified by substituting the parameter $M$ with a related 
parameter, $m$, that is defined by (Equation 5 of M24)
\begin{equation}
m = \frac{\langle X_A\rangle - \langle X_B\rangle}{\langle X_A\rangle + \langle X_B\rangle}
  = \frac{M-1}{M+1}, \qquad -1\le m\le 1.
\label{eqn:m}
\end{equation}
The emission is composed solely of mode A when $m=1$ ($M=\infty$) and is composed solely 
of mode B when $m=-1$ ($M=0$). Changes in $m$ with pulse longitude drive the transition 
between modes. If the polarization mode vectors are strictly orthogonal, the transition 
between modes occurs at $m=0$ ($M=1$). 

 
\subsection{Partially Coherent OPMs}
\label{sec:pcoh}

OKJ's implementation of the PCOH model assumes the radio emission is composed of two 
partially coherent, linearly polarized OPMs. Their Equation 5 shows that the model is 
characterized by three parameters: a mode coherence fraction, $C$, a mode phase offset, 
$\eta$, and a mode strength ratio, $R$, which is identical to $M$. The coherence fraction 
ranges from $C=0$ for incoherent OPMs to $C=1$ for completely coherent OPMs. The mode 
phase offset must vary over a range of $0\le\eta\le 2\pi$ for a vector rotation to 
complete an SC. After simplifying Equations 36 and 39 of M24, the equations for the EA 
and polarization fraction derived from the PCOH model can be shown to be 
\begin{equation}
\sin^2(2\chi) = \frac{\sin^2(\eta)(1-m^2)}{m^2(t^2 - 1) + 1},
\label{eqn:EAP}
\end{equation}
\begin{equation}
p^2 = \frac{m^2(t^2-1) + 1}{t^2}.
\label{eqn:pp}
\end{equation}
The parameter $t$ in Equations~\ref{eqn:EAP} and~\ref{eqn:pp} is a function of the coherence 
fraction and is given by (M24)
\begin{equation}
t = \frac{(1-C)^2+C^2}{C^2}.
\label{eqn:t}
\end{equation}
The value of $t$ ranges from $t=1$ when $C=1$ to $t=\infty$ when $C=0$. 

The PCOH model draws a clear distinction between variations in $p$ and $\chi$ due to a mode 
transition versus those arising from a polarization vector rotation. The overall process of 
an ideal mode transition occurs when $m$ varies, while $C$ and $\eta$ are constant. The actual 
transition between modes occurs at $m=0$, where the EA attains its peak value of $\chi=\eta/2$, 
and the polarization fraction is minimum, $p_m=1/t$. Both $\chi$ and $p$ vary symmetrically 
about $m=0$ (i.e., $\chi(m)=\chi(-m)$ and $p(m)=p(-m)$; M24). An ideal vector rotation occurs 
when $\eta$ varies while $m$ and $C$ are constant. The polarization fraction remains constant 
over the rotation, because it is independent of $\eta$. The peak value of $|\chi|$ in a vector
rotation occurs at $\eta=\pm\pi/2$. 

\begin{figure}
\plotone{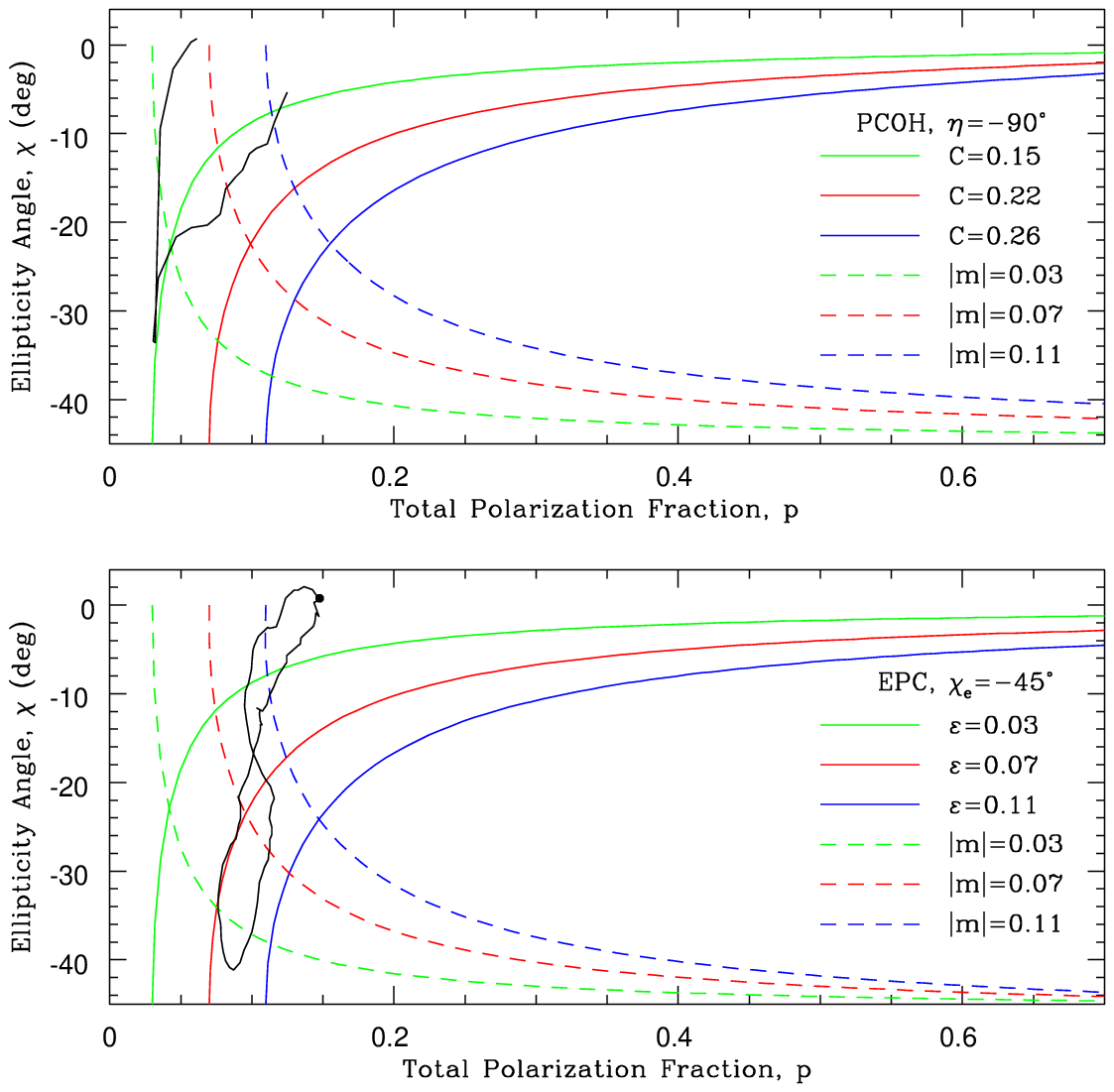}
\caption{Variations in the polarization fraction and EA due to changes in the parameters of
the PCOH and EPC polarization models. Top panel: the solid colored lines show the variations 
in $p$ and $\chi$ expected from the PCOH model when the parameter $m$ varies, while $\eta$
and $C$ are held constant. The dashed lines show the $p$-$\chi$ variations caused by changes 
in $C$ while $\eta$ and $|m|$ are held constant. The solid black line shows the $p$-$\chi$ 
variations observed in PSR B1133+16. Bottom panel: the solid colored lines show the variations 
in $p$ and $\chi$ expected from the EPC model when the parameter $m$ varies, while $\chi_e$ 
and $\varepsilon$ are held constant. The dashed lines show the $p$-$\chi$ variations caused 
by changes in $\varepsilon$ while $\chi_e$ and $|m|$ are held constant. The solid black line 
shows the $p$-$\chi$ variations observed in PSR B2016+28.}
\label{fig:models}
\end{figure}

The top panel of Figure~\ref{fig:models} is a variant of the diagram used by OKJ to illustrate 
their PCOH model (see their Figure 1). It shows how $p$ and $\chi$ change with variations in 
the model parameters, whereas OKJ's Figure 1 illustrates changes in $p$ and the absolute value 
of the polarization vector's latitude, $\theta$. The solid lines in the panel show ideal mode 
transitions for three different values of $C$ with the phase offset fixed at $\eta=-90^\circ$. 
Since a mode transition in the PCOH model is symmetric about $m=0$, the $p$-$\chi$ track for 
$m=0$ to $m=1$ is the same as the track for $m=0$ to $m=-1$. Therefore, the solid lines are 
drawn for changes in the absolute value of $m$. It increases along each line from $|m|=0$ in 
the lower-left corner of the panel toward $|m|=1$ in the upper-right corner. The minimum 
values of the polarization fraction ($p_m=1/t$) and the EA ($\chi=\eta/2$) occur at $m=0$. 
As $|m|$ initially increases from zero, the EA increases significantly while the polarization 
fraction remains nearly constant at $p_m$. But as $|m|$ increases further, it is the 
polarization fraction that increases substantially while the EA increases only gradually. At 
$|m|=1$, the polarization fraction is $p=1$, and the EA is $\chi=0$. The dashed lines in the 
panel show how $p$ and $\chi$ vary with changes in $C$, while the parameters $|m|$ and $\eta$ 
are held constant. The coherence fraction increases along each dashed line from $C=0$ in the 
upper-left corner of the panel toward $C=1$ in the lower-right corner. The EA is $\chi=0$ 
and the polarization fraction is minimum at $p_m=|m|$ when the modes are incoherent ($C=0$). 
As $C$ initially increases from zero, the EA abruptly decreases while the polarization 
fraction remains roughly constant near its minimum value. As the modes become increasingly 
coherent, the polarization fraction increases, and the EA gradually approaches $\chi=\eta/2$. 
An ideal vector rotation is not shown in the panel. It would appear as a vertical line 
located at a value of $p$ determined by the fixed values of $C$ and $m$ (see Figure 1 of 
OKJ). The change in EA (i.e., the length of the vertical line) would be determined by the 
total change in $\eta$. The solid black line in the panel shows the observed $p$-$\chi$ 
track covering the first PA discontinuity in PSR B1133+16. It is discussed in 
Section~\ref{sec:1133}.


\subsection{Incoherent OPMs with an Elliptically Polarized Emission Component}

The EPC model of pulsar polarization assumes the radio emission is composed of two incoherent, 
linearly polarized OPMs and an independent, elliptically polarized emission component (S84; 
ES04; M24). The model is characterized by three parameters: the relative intensity and the EA 
of the EPC, $\varepsilon$ and $\chi_e$, respectively, and $m$. The EPC intensity lies in the 
range $0\le\varepsilon\le 1$, and the range of the EPC EA is $-\pi/4\le\chi_e\le\pi/4$. The EA 
and polarization fraction derived from the EPC model are (Equations 28 and 33, respectively, 
of M24)
\begin{equation}
\sin^2(2\chi) = \frac{\varepsilon^2\sin^2(2\chi_e)}{m^2(1-\varepsilon)^2 + \varepsilon^2},
\label{eqn:EAE}
\end{equation}
\begin{equation}
p^2 = m^2(1-\varepsilon)^2 + \varepsilon^2.
\label{eqn:pe}
\end{equation}

The EPC model is similar, but not identical, to the PCOH model. An ideal mode transition 
in the EPC model occurs when $m$ varies, while $\varepsilon$ and $\chi_e$ are constant. 
The actual mode transition occurs at $m=0$, where the observed EA attains its peak value
of $\chi=\chi_e$, and the polarization fraction is minimum, $p_m=\varepsilon$. As with the 
PCOH model, both $\chi$ and $p$ vary symmetrically about $m=0$. An ideal vector rotation 
occurs when $\chi_e$ varies while $m$ and $\varepsilon$ are constant. The polarization 
fraction remains constant over the rotation, because it is independent of $\chi_e$. The 
peak value of $|\chi|$ occurs at the largest value of $|\chi_e|$ in the rotation. 

The bottom panel of Figure~\ref{fig:models} illustrates how $p$ and $\chi$ vary with changes 
in the parameters of the EPC model. The solid lines show ideal mode transitions for three 
different values of $\varepsilon$ with the EPC EA fixed at $\chi_e=-45^\circ$. They are very
similar to the $p$-$\chi$ tracks of the mode transitions in the PCOH model. The absolute 
value of $m$ increases along each solid line from $|m|=0$ in the lower-left corner of the 
panel toward $|m|=1$ in the upper-right corner. The minimum values of the polarization 
fraction ($p_m=\varepsilon$) and the EA ($\chi=\chi_e$) occur at $|m|=0$. The dashed lines 
in the panel show how $p$ and $\chi$ vary with changes in $\varepsilon$ while the 
parameters $|m|$ and $\chi_e$ are held constant. The EPC intensity increases along each 
line from $\varepsilon=0$ in the upper-left corner of the panel toward $\varepsilon=1$ 
in the lower-right corner. When the EPC is not present ($\varepsilon=0$), the EA is 
$\chi=0$, and the minimum polarization fraction is $p_m=|m|$. As the relative intensity of 
the EPC initially increases from zero, the EA abruptly decreases while the polarization 
fraction remains roughly constant near its minimum value. The polarization fraction then 
increases, and the EA gradually converges toward $\chi=\chi_e$ as $\varepsilon$ increases 
toward 1. An ideal vector rotation is not shown in the panel. It would appear as a 
vertical line located at a value of $p$ determined by the fixed values of $\varepsilon$ 
and $m$. The length of the vertical line would be determined by the total change in 
$\chi_e$. The solid black line in the panel shows the observed $p$-$\chi$ track covering 
most of the pulse of PSR B2016+28. It is discussed in Section~\ref{sec:2016}.


\subsection{Nonorthogonal Polarization Modes}

The NPM model of pulsar polarization assumes the radio emission is composed of two incoherent 
polarization modes whose polarization vectors are not precisely orthogonal (M24). The three 
parameters that characterize the model are $m$ and the departures from orthogonality in linear 
polarization, $\delta_l$, and circular polarization, $\delta_v$. While the model can 
accommodate departure angles in the range $0\le|\delta|\le\pi/4$, it implicitly assumes the 
angles are small, $|\delta|\ll 1$. The model is tailored to mode transitions. 

The EA and polarization fraction derived from the NPM model are (Equations 19 and 20 of M24)
\begin{equation}
\sin^2(2\chi) = \frac{\sin^2(2\delta_v)(1-m)^2}{2(K_1 + m^2K_2)},
\label{eqn:EAgen}
\end{equation}
\begin{equation}
p^2 = \frac{K_1 + m^2K_2}{2},
\label{eqn:pn}
\end{equation}
where $K_1$ and $K_2$ are given by $K_1=1-\cos(2\delta_v)\cos(2\delta_l)$ and
$K_2=1+\cos(2\delta_v)\cos(2\delta_l)$. An ideal mode transition in the NPM model occurs 
when $m$ varies, while $\delta_l$ and $\delta_v$ remain constant. The polarization fraction 
of a mode transition is symmetric about $m=0$, but the EA is not. The transition occurs at 
$m = -K_1/K_2$, where the polarization fraction is $p=\sqrt{K_1/K_2}$ and the magnitude 
of the EA is maximum. The minimum polarization fraction occurs at $m=0$ and is equal to 
$p_m = \sqrt{K_1/2}$. When the polarization modes are orthogonal ($\delta_l=\delta_v=0$), 
the polarization fraction is $p=|m|$ and the EA is $\chi=0$. 

\begin{figure}
\plotone{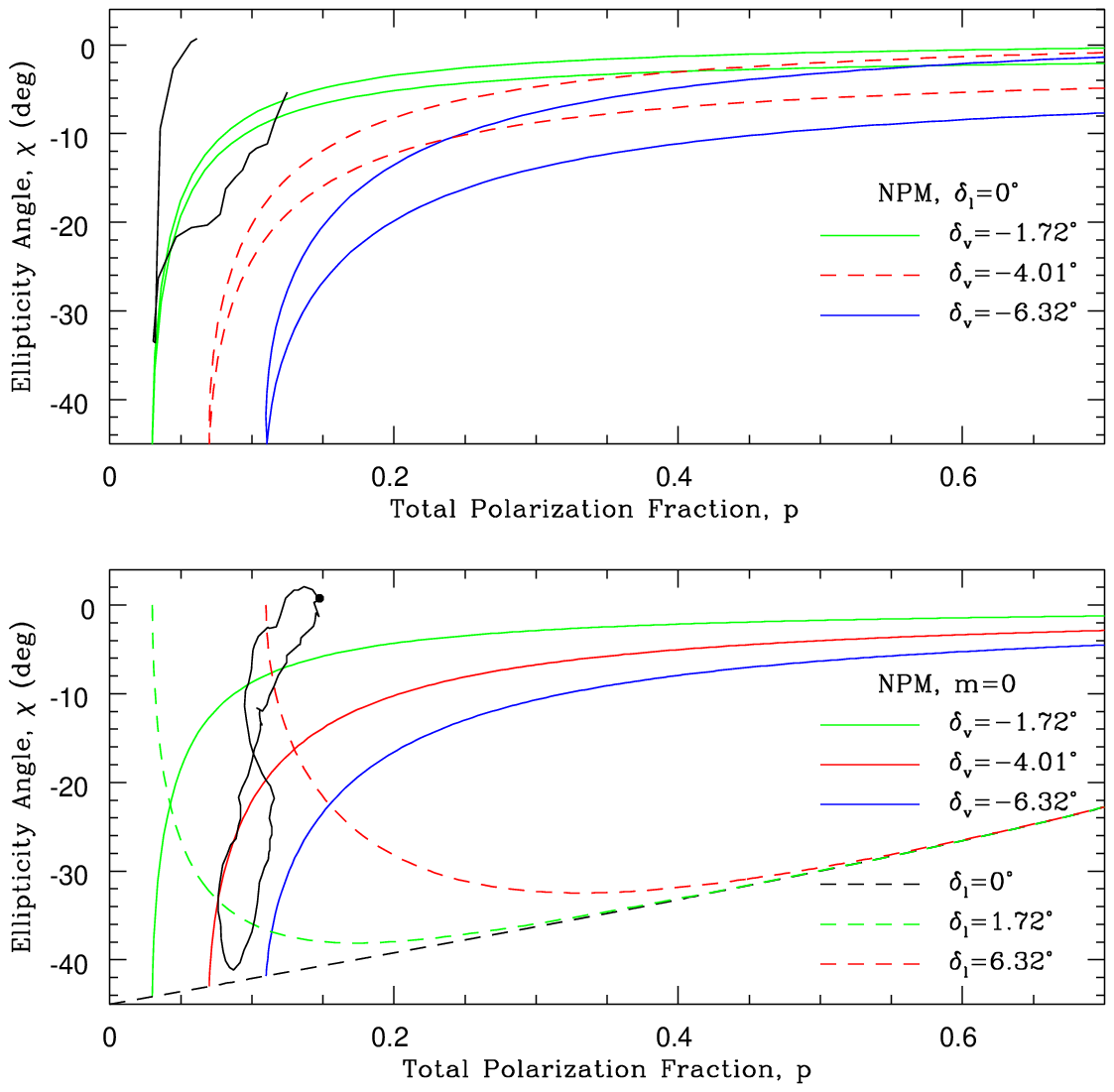}
\caption{Variations in the polarization fraction and EA due to changes in the parameters 
of the NPM polarization model. Top panel: the colored lines show the expected variations 
in $p$ and $\chi$ when $m$ varies, while $\delta_l$ and $\delta_v$ are held constant. The 
solid black line shows the $p$-$\chi$ variations observed in PSR B1133+16. Bottom panel: 
the solid colored lines show the expected variations in $p$ and $\chi$ when $\delta_l$ 
varies, while $m$ and $\delta_v$ are held constant. The dashed lines show the $p$-$\chi$ 
variations caused by changes in $\delta_v$, while $m$ and $\delta_l$ are held constant. 
The solid black line shows the $p$-$\chi$ variations observed in PSR B2016+28.}
\label{fig:npmmodel}
\end{figure}

Figure~\ref{fig:npmmodel} illustrates how $p$ and $\chi$ vary with changes in the model
parameters. Each set of lines in the top panel of the figure shows an ideal mode transition 
for three different values of $\delta_v$ with $\delta_l$ fixed at zero. The values of 
$\delta_v$ listed in the panel were selected to reproduce the same values of $p$ near 
$\chi=-\pi/4$ as in Figure~\ref{fig:models}. Unlike the PCOH and EPC models, a mode 
transition in the NPM model is not symmetric about $m=0$. Therefore, the $p$-$\chi$ track 
for $m=0$ to $m=1$ is different from the track for $m=0$ to $m=-1$. The top line in each 
set of lines shows the track from $m=1$ near the top-right corner of the panel to $m=0$ in 
the bottom-left corner. The bottom line of each set shows the remainder of the track from 
$m=0$ to $m=-1$ near the top-right corner. The EA and polarization fraction at $m=1$ are 
$\chi=0$ and $p=1$. As $m$ decreases from 1, the polarization fraction decreases 
significantly, while the EA decreases only gradually. As $m$ approaches zero, the EA 
abruptly decreases toward $\chi=|\delta_v|/2-\pi/4$, and the polarization fraction remains 
roughly constant near its minimum value of $p_m=|\sin(\delta_v)|$. The magnitude of the EA 
attains its largest value of $|\chi|=\pi/4$ at $m=-\tan^2(\delta_v)$, where the polarization 
fraction has increased slightly to $p=|\tan(\delta_v)|$. As $m$ becomes increasingly negative, 
the EA initially increases significantly at a roughly constant polarization fraction, but then 
increases only gradually, while $p$ increases substantially as $m$ approaches $m=-1$. The EA 
and polarization fraction at $m=-1$ are $\chi=\delta_v$ and $p=1$.

The bottom panel of the figure illustrates how $\chi$ and $p$ vary with $\delta_v$ and 
$\delta_l$ when $m$ is held constant at zero. The solid lines are drawn for fixed values
of $\delta_v$ as $\delta_l$ varies from $\delta_l=0$ in the bottom-left corner of the panel
toward $\delta_l=\pi/4$ near the top-right corner. The $p$-$\chi$ tracks in this instance
resemble the mode transition tracks of the PCOH and EPC models. When $\delta_l=0$, the 
relationship between the EA and $\delta_v$ is $\sin(2\chi)=-\cos(\delta_v)$, and the 
polarization fraction is $p=|\sin(\delta_v)|$. Thus, the lower bound on the EA at 
$\delta_l=0$ is $\chi=-\arccos(p)/2$, as indicated by the dashed black line in the panel. 
The dashed lines are drawn for fixed values of $\delta_l$ as $\delta_v$ varies from 
$\delta_v=0$ on the left side of the panel toward $\delta_v=-\pi/4$ on the right side. 
For all values of $\delta_l$ except zero, the EA initially decreases abruptly with 
increasing $\delta_v$ while the polarization fraction remains roughly constant near its 
minimum value of $p_m=\sin(\delta_l)$. As $\delta_v$ approaches $-\pi/4$, the EA converges 
to $\chi=-\pi/8$, and the polarization fraction converges to $p=1/\sqrt{2}$ for all values 
of $\delta_l$.


\section{Model Comparison with Observations of PSR B1133+16}
\label{sec:1133}

The solid black line in the top panels of Figures~\ref{fig:models} and~\ref{fig:npmmodel}
shows the observed $p$-$\chi$ track covering the first PA discontinuity in PSR B1133+16. 
The track consists of two branches. The first branch begins at ppb 92, where the EA is
$-5.3^\circ$ and the polarization fraction is $p=0.125$. The track initially descends 
gradually to the left and then abruptly to $\chi=-33.6^\circ$ and $p=0.032$ at ppb 107.
This branch passes through the minimum polarization fraction of $p_m=0.031$ at ppb 106, 
where $\chi=-33.4^\circ$. The second branch of the track begins in the vicinity of 
bins 106-107 and ascends almost vertically to $\chi=0.7^\circ$ and $p=0.061$ at ppb 112.
The overall track is inconsistent with an ideal vector rotation, because it does not 
trace a single vertical line. The track is also inconsistent with an ideal mode 
transition in the PCOH and EPC models, because its two branches do not follow the same 
path in $p$-$\chi$ space. The two branches bear some resemblance to an ideal mode 
transition in the NPM model.

The inconsistencies noted here do not necessarily mean that the PCOH and EPC models are 
not applicable to the observation. The inconsistencies could arise from the assumption
that only one model parameter varies in a mode transition or a vector rotation. The
fact that some observed PA-EA pairs do not reside precisely on the GC drawn in the top 
panel of Figure~\ref{fig:EAPA} suggests that more than one parameter may vary over the 
GC trajectory.

The parameters that characterize the PCOH, EPC, and NPM models can be calculated from 
the observed polarization fraction and EA by inverting the equations for $p$ and $\chi$ 
derived for each model. The equations for the parameters written as functions of $p$
and $\chi$ are listed in Appendix~\ref{sec:parms}. Since each model contains three 
parameters, and there are only two observable quantities, the value of one parameter 
must be estimated to solve for the remaining two. The estimated parameter can be 
constrained by the observations, as demonstrated in Appendix~\ref{sec:parms}. Three 
different cases must be evaluated for each model, because there are three different 
combinations of the parameters when one of them is assumed to be fixed. For 
convenience, each case is nominally designated as one of a mode transition, a vector 
rotation, or a transition-rotation hybrid, but with the caveat that these 
designations may not accurately describe some cases of the NPM model.


\subsection{PCOH model}


\subsubsection{Mode Transition}

\begin{figure}
\plotone{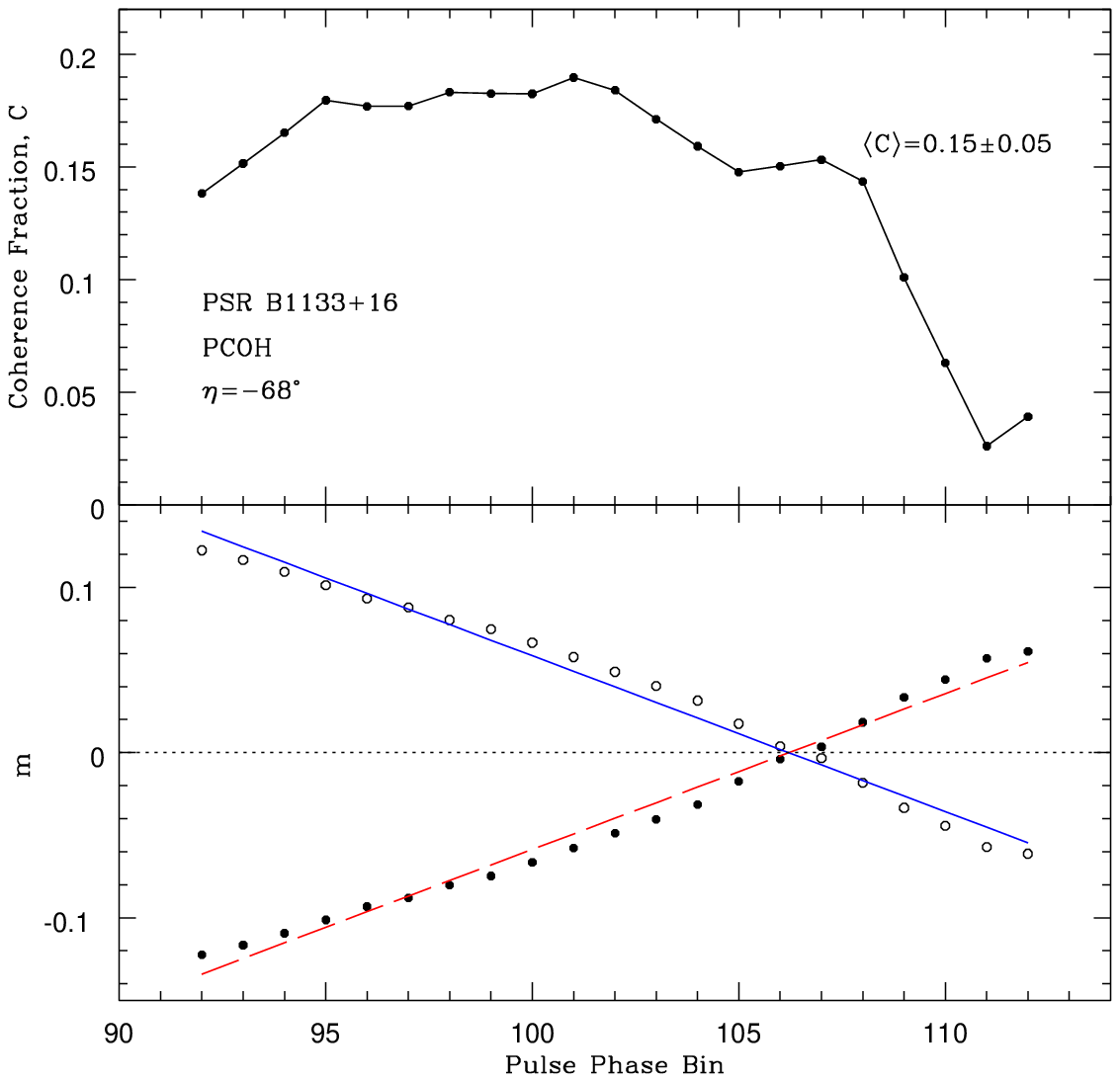}
\caption{Results of the PCOH model applied to the first PA discontinuity in PSR B1133+16 
assuming the observed variations in $p$ and $\chi$ are caused by a mode transition with 
a varying coherence fraction. The top panel shows how the calculated coherence fraction 
varies with ppb. The circles in the bottom panel show how the calculated values of $m$ 
vary with ppb. The lines represent the best fit of the data points to a straight line. 
The mode phase offset used in the calculation was $\eta=-68^\circ$.}
\label{fig:pcoh1133}
\end{figure}

The first case to consider with the PCOH model is a mode transition where the parameter 
$m$ and the coherence fraction are allowed to vary while the mode phase offset remains 
fixed. The values of the parameters $m$ and $C$ at each ppb can be calculated from the 
observed values of $p$ and $\chi$ using Equations~\ref{eqn:cm}, ~\ref{eqn:ct}, 
and~\ref{eqn:t}. From Appendix~\ref{sec:parms}, the fixed value of $|\eta|$ is 
constrained to be greater than or equal to twice the largest value of $|\chi|$ within 
the pulse region of interest. For PSR B1133+16, this constraint gives $|\eta|\ge 68^\circ$ 
and occurs at ppb 107. The results of the mode transition calculation are shown in 
Figure~\ref{fig:pcoh1133}. The top panel of the figure shows the calculated coherence 
fraction hovers near $C\simeq 0.17$ on the leading edge of the PA discontinuity but drops 
to near-zero on the trailing edge. Overall, the mean value of $C$ in this part of the 
pulse is $\langle C\rangle = 0.15\pm 0.05$. The low coherence fraction is expected given 
that the polarization fraction observed across the pulse is low. The open and filled 
circles in the bottom panel of the figure show that the parameter $m$ varies linearly 
with ppb. Two sets of solutions for $m$ are allowed, because $m$ is symmetric about $m=0$ 
in the model. This also means it is not possible to determine whether $m$ is increasing 
or decreasing with ppb (OKJ). The two lines in the panel represent best fits of straight 
lines to the data points. Both lines cross $m=0$ near ppb 106. The slopes, intercepts,
and correlation coefficients of the lines are listed in Table 1. The number of 
significant digits used in the table entries is intended to illustrate the subtle 
differences between the models' fit parameters and not to infer a level of statistical 
significance.

\subsubsection{Vector Rotation}

The second case to consider is a vector rotation where the mode phase offset and the 
coherence fraction are allowed to vary while the parameter $m$ remains fixed. The values 
of $\eta$ and $C$ at each ppb can be calculated from the observed values of $p$ and 
$\chi$ using Equations~\ref{eqn:eta}, ~\ref{eqn:tm}, and~\ref{eqn:t}. From 
Appendix~\ref{sec:parms}, the fixed value of $m$ is constrained by $p\cos(2\chi)\ge 
|m|$ everywhere within the pulse region of interest. This constraint gives $|m|\le 0.012$ 
and occurs at ppb 106. The results of the vector rotation calculation are not shown. 
The calculated value of $C$ is low and variable with a mean value of 
$\langle C\rangle = 0.21\pm 0.04$. The coherence fraction is minimum at ppb 106 where 
the polarization fraction is also minimum. This is expected, because the only way to 
decrease $p$ when $m$ is constant is through a loss of coherence. The calculated value 
of $\eta$ varies across this region of the pulse, but not in the organized, linear 
fashion as the parameter $m$ in the mode transition case. 

Overall, the vector rotation of the PCOH model is not consistent with the observation of 
PSR B1133+16, because the model's implementation requires the plane of the SC or GC to be 
perpendicular to the equatorial plane of the Poincar\'e sphere (see Figure~\ref{fig:gcsc}), 
whereas the observed GC is inclined to the plane. An alternate interpretation is the 
OPMs are elliptically polarized, instead of linearly polarized, as assumed in the model's
implementation (see Equation 5 of OKJ). In this instance, the intrinsic EA of the modes 
is then the EA of the normal vector to the GC plane, $\chi_n=11^\circ$ (see 
Figure~\ref{fig:EAPA}). Furthermore, (i) the value of $m$ must be zero for the PA-EA pairs 
to trace a GC, (ii) the coherence fraction can be determined from $t=1/p$, and (iii) the 
mode phase offset can be calculated with 
\begin{equation}
\sin(\eta)=\sin(2\chi)/\cos(2\chi_n).
\label{eqn:etaell}
\end{equation}
The results returned by this interpretation are almost identical to those of the original 
vector rotation calculation. 


\subsubsection{Transition-rotation Hybrid}

The third case to consider is a transition-rotation hybrid where the parameter $m$ and 
the mode phase offset are allowed to vary while the coherence fraction remains fixed. 
The values of $m$ and $\eta$ at each ppb can be calculated from the observed values of 
$p$ and $\chi$ using Equations~\ref{eqn:hm} and~\ref{eqn:heta}. As described in 
Appendix~\ref{sec:parms}, the fixed value of $t$ is constrained by upper and lower 
limits. The lower limit is $t\ge 1/p_{m}$, where $p_m$ is the minimum polarization 
fraction in the pulse region of interest. This constraint occurs at ppb 106, where $t=33$, 
or $C=0.15$. The upper limit on $t$ is given by Equation~\ref{eqn:tmax}. It occurs at 
ppb 101, where $t=21$, or $C=0.18$. The two limits are mutually exclusive, because the
lower limit on $t$ exceeds its upper limit. Consequently, solutions for both $m$ and 
$\eta$ cannot be determined at all ppbs. Therefore, the transition-rotation hybrid of 
the PCOH model is not a viable interpretation of the observation. 

\subsection{EPC Model}


\subsubsection{Mode Transition}

For a mode transition in the EPC model, the parameter $m$ and the EPC relative intensity 
vary, while the EPC EA remains fixed. The values of $m$ and $\varepsilon$ at each ppb can 
be calculated from $p$ and $\chi$ using Equations~\ref{eqn:me} and~\ref{eqn:ve}. The 
fixed value of $|\chi_e|$ is constrained by the largest value of $|\chi|$ within the 
pulse region of interest. This constraint gives $\chi_e=-34^\circ$ and occurs at ppb 107. 
The results of the mode transition calculation are not shown. The calculation shows the 
parameter $m$ varies linearly with ppb. A straight-line fit of the calculated values of 
$m$ yields fit parameters that are very similar to those obtained with a mode transition 
in the PCOH model. The fit parameters are listed in Table 1. The calculated values of 
$\varepsilon$ are low and variable with a mean value of 
$\langle\varepsilon\rangle = 0.03\pm 0.02$. 

\begin{deluxetable}{cccc}
\tablenum{1}
\tablecaption{Linear Variation of the Parameter $m$ with Pulse Phase Bin in PSR
              B1133+16}
\tablehead{\colhead{Model} & \colhead{PCOH} & \colhead{EPC} & \colhead{NPM}}
\startdata
Constant Term & $\eta=-68^\circ$ & $\chi_e=-34^\circ$ & $\delta_l=0.57^\circ$ \\
Variable Term & $\langle C\rangle = 0.15\pm 0.05$ & $\langle\varepsilon\rangle 
              = 0.03\pm 0.02$ & $\langle\delta_v\rangle=-1.8^\circ\pm 0.9^\circ$ \\
Slope ($\times 10^3$) & $\pm 9.43$ & $\pm 9.71$ & $+9.48, -9.35$ \\
Intercept & $\mp 1.00$ & $\mp 1.03$ & $-1.01, +0.99$ \\
Corr. Coeff. & $\pm 0.992$ & $\pm 0.992$ & $+0.990, -0.992$ \\
\enddata
\end{deluxetable}


\subsubsection{Vector Rotation}

A vector rotation in the EPC model fixes the value of $m$ while the relative intensity 
and EA of the EPC are allowed to vary. The values of $\varepsilon$ and $\chi_e$ at each 
ppb can be calculated from $p$ and $\chi$ using Equations~\ref{eqn:Iep} and~\ref{eqn:EAp}. 
The fixed value of $m$ is constrained by $|m|\le p_m$ everywhere within the pulse region 
of interest. This constraint gives $|m|\le 0.03$ and occurs at ppb 106. The results of 
the vector rotation calculation are not shown. The calculated value of $\varepsilon$ is 
low and variable with a mean value of $\langle\varepsilon\rangle = 0.07\pm 0.03$. The
calculated values of $\chi_e$ closely track the observed EA.


\subsubsection{Transition-rotation Hybrid}

In a transition-rotation hybrid of the EPC model, the parameter $m$ and the EPC EA vary,
while the EPC intensity remains fixed. The values of $\chi_e$ and $m$ at each ppb can 
be calculated from $p$ and $\chi$ using Equations~\ref{eqn:EAp} and~\ref{eqn:mepc}. The 
fixed value of $\varepsilon$ is constrained by an upper limit of $\varepsilon\le p_{m}$
and a lower limit of $\varepsilon\ge p|\sin(2\chi)|$. The upper limit occurs at ppb 106,
where $p_m=\varepsilon=0.03$. The lower limit occurs at ppb 101, where $\varepsilon=0.05$. 
The two limits are mutually exclusive, because the lower limit on $\varepsilon$ exceeds 
its upper limit. Consequently, solutions for both $m$ and $\chi_e$ cannot be determined 
at all ppbs. Therefore, the transition-rotation hybrid of the EPC model is not a viable 
interpretation of the observation. 

\subsection{NPM Model}


\subsubsection{Mode Transition}

\begin{figure}
\plotone{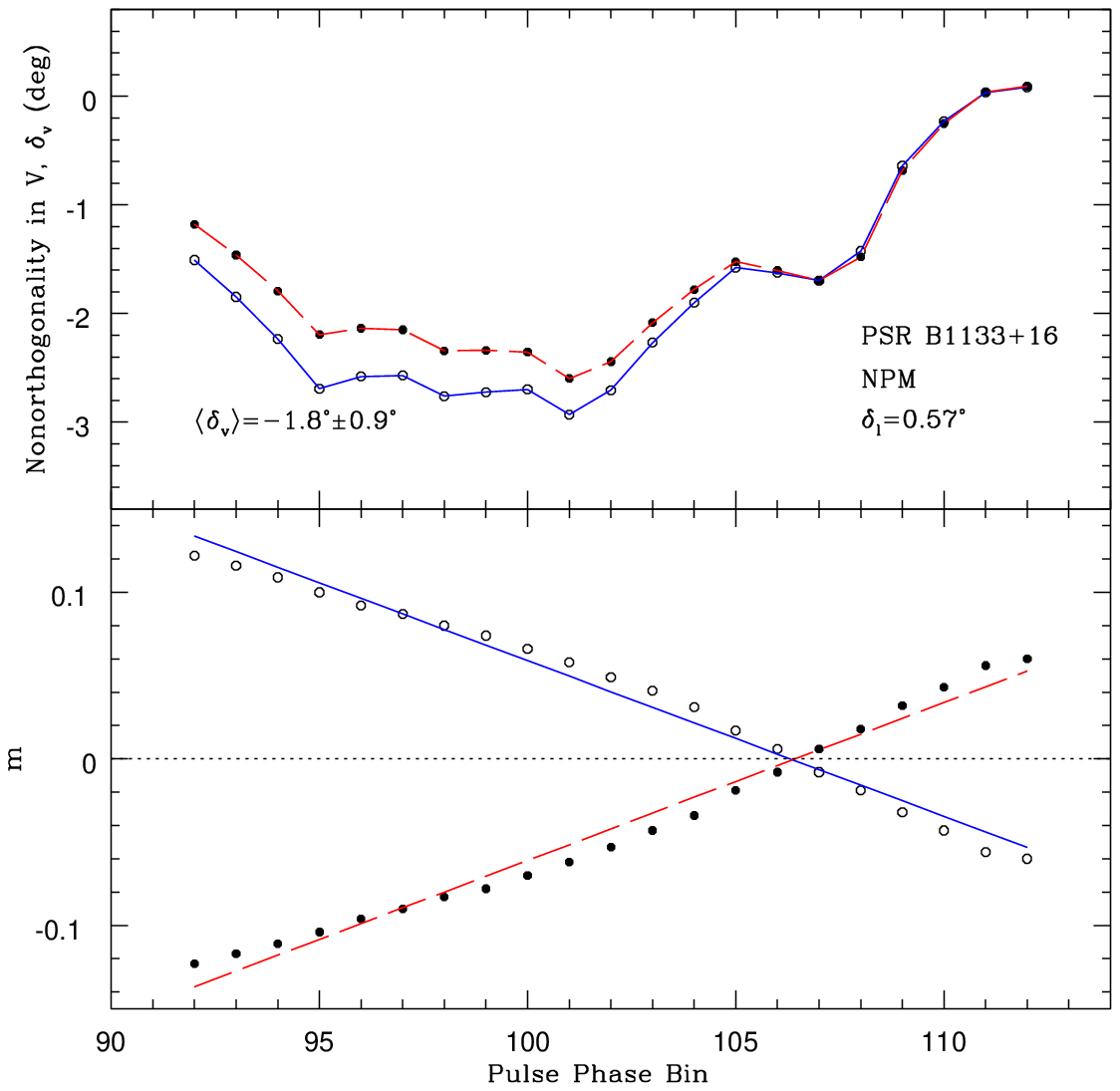}
\caption{Results of the NPM model applied to PSR B1133+16 assuming the observed variations 
in $p$ and $\chi$ are caused by a mode transition with varying nonorthogonality in circular 
polarization, $\delta_v$. The parameter $m$ is shown in the bottom panel, and the value of 
$\delta_v$ is shown in the top panel. The values of $\delta_v$ and $m$ that follow the 
dashed red lines are associated with one another. The same is true of the values of $\delta_v$
and $m$ that follow the solid blue lines. The nonorthogonality in linear polarization used
in the calculation was held constant at $\delta_l=0.57^\circ$.}
\label{fig:npm1133mt}
\end{figure}

A mode transition occurs in the NPM model when the parameters $m$ and $\delta_v$ vary,
while $\delta_l$ remains fixed. The values of $m$ and $\delta_v$ at each ppb can be 
calculated from the observed values of $p$ and $\chi$ using Equations~\ref{eqn:mn} 
and~\ref{eqn:deltav}. Two values of each of $m$ and $\delta_v$ are allowed at each ppb,
because they are solutions to quadratic equations and depend upon whether $m$ increases 
or decreases with pulse longitude (see Appendix~\ref{sec:parms}). The fixed value of 
$\delta_l$ is constrained by $|\sin(\delta_l)|\le p\cos(2\chi)$ everywhere within the 
pulse region of interest. This constraint gives $\delta_l=0.57^\circ$ and occurs at ppb 
106. The results of the mode transition calculation are shown in Figure~\ref{fig:npm1133mt}. 
The bottom panel of the figure shows that $m$ varies linearly with ppb. A straight-line 
fit of the calculated values of $m$ yields fit parameters that are similar to those 
obtained with a mode transition in the PCOH and EPC models (Table 1). However, the 
magnitudes of the fit parameters for the two lines are not identical, because an NPM mode 
transition is not symmetric about $m=0$. The top panel shows that the calculated values 
of $\delta_v$ are small and variable. The values of $\delta_v$ and $m$ that follow the 
dashed red lines are associated with one another, as are the values of $\delta_v$ and 
$m$ that follow the solid blue lines. The mean value of 
$\delta_v$, $\langle\delta_v\rangle=-1.8^\circ\pm 0.9^\circ$, annotated in the panel 
was calculated from the data points following the solid blue line. The small values of 
$\delta_l$ and $\delta_v$ returned by the analysis are consistent with the NPM model's 
assumption that the departures from orthogonality are small. The analysis results also 
demonstrate that small departures from orthogonality can produce large changes in the 
observed EA.


\subsubsection{Pseudo-vector Rotation}

\begin{figure}
\plotone{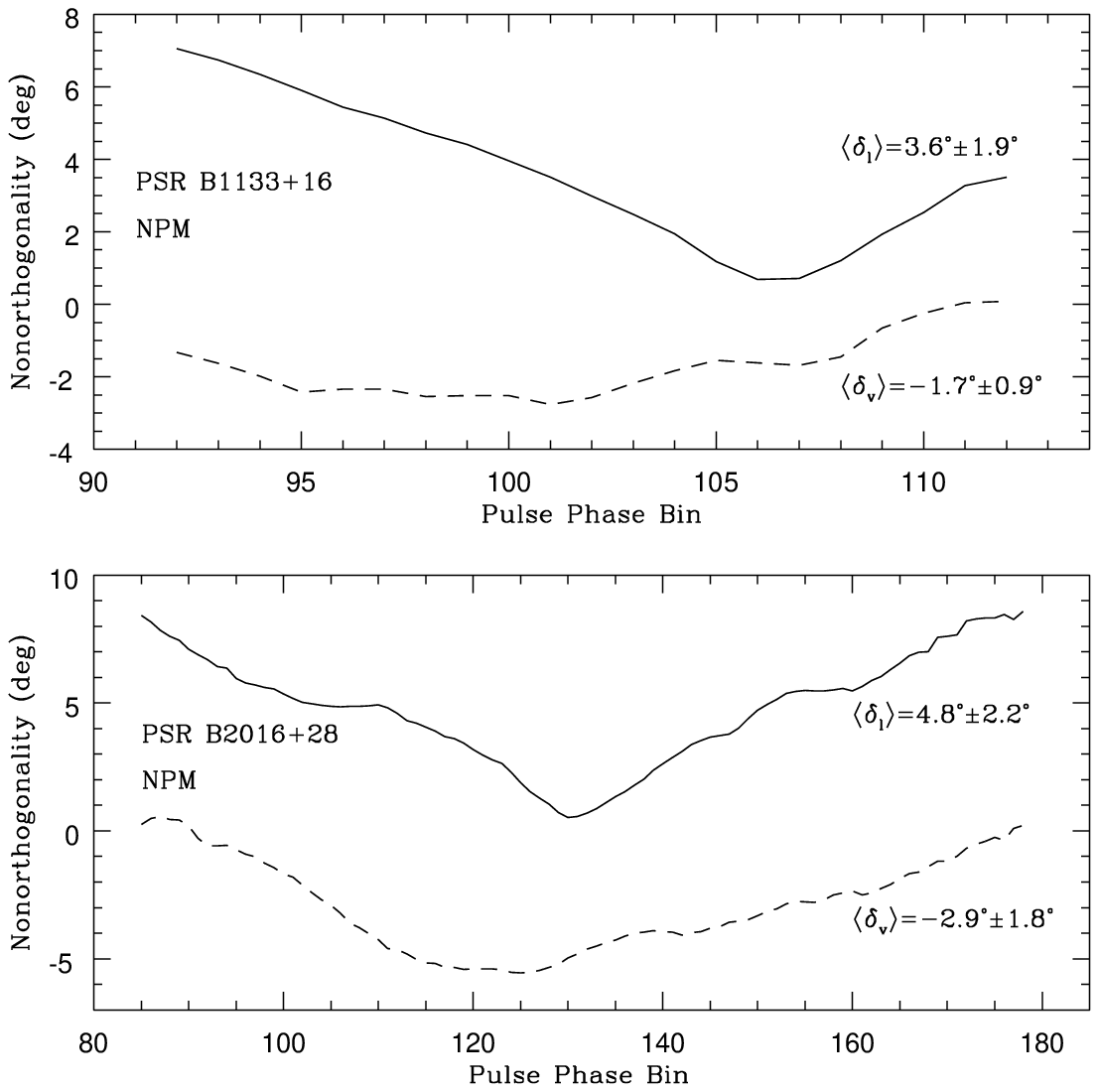}
\caption{Results of the NPM model applied to PSR B1133+16 (top panel) and PSR B2016+28 (bottom
panel) assuming the observed variations in $p$ and $\chi$ are caused by variations in the 
parameters $\delta_l$ and $\delta_v$. The calculated nonorthogonality in linear polarization 
is represented by the solid line, and the nonorthogonality in circular polarization is denoted 
by the dashed line. The parameter $m$ was set to zero in the calculations.}
\label{fig:npm0}
\end{figure}

The label of \lq\lq pseudo-vector rotation" is assigned to the case of the NPM model, where 
$\delta_l$ and $\delta_v$ vary, while $m$ remains fixed. The case is analogous to the vector 
rotation cases of the PCOH and EPC models, where $m$ is fixed while the other model parameters 
vary. The \lq\lq pseudo" modifier is included in the label to denote that the variations
in $\delta_l$ and $\delta_v$ are not necessarily true rotations. The values of $\delta_l$ 
and $\delta_v$ at each ppb can be calculated from the observed values of $p$ and $\chi$ 
using Equations~\ref{eqn:lrot} and~\ref{eqn:vrot}. The fixed value of $m$ is constrained by 
$|m|\le p_m$ everywhere within the pulse region of interest. The results of the calculation 
using $m=0$ are shown in the top panel of Figure~\ref{fig:npm0}. The calculated values of 
$\delta_l$ and $\delta_v$ are small, as implicitly assumed in the model. The values of 
$\delta_v$ resemble those determined in the mode transition case. The mean of $\delta_v$ is 
$\langle\delta_v\rangle=-1.7^\circ\pm 0.9^\circ$ and is consistent with 
$\langle\delta_v\rangle$ calculated for the mode transition. The value of $\delta_l$ reaches 
a minimum, and is less than $|\delta_v|$, in the vicinity of the EA maximum near ppb 106-107.


\subsubsection{Pseudo-transition-rotation Hybrid}

The label of \lq\lq pseudo-transition-rotation hybrid" is assigned to the case of the NPM 
model where $m$ and $\delta_l$ vary, while $\delta_v$ remains fixed. The transition aspect 
of the hybrid case is caused by variations in $m$, while the rotation aspect is caused by 
variations in $\delta_l$. The values of $m$ and $\delta_l$ at each ppb can be calculated 
from the observed values of $p$ and $\chi$ using Equations~\ref{eqn:mn} and~\ref{eqn:deltal}. 
From Appendix~\ref{sec:parms}, the fixed value of $\delta_v$ is constrained by 
$\tan(\delta_v)\le p\sin(2\chi)$ everywhere within the pulse region of interest. The 
constraint occurs at ppb 101 where $\delta_v=-2.76^\circ$. It is difficult for this 
interpretation to produce large changes in $\chi$ by altering $\delta_l$. The calculation 
produces solutions for $m$, but they generally exceed their respective values of $p$, which 
is inconsistent with the requirement that $|m|\le p$ (see Equation~\ref{eqn:pn}). Also, the 
calculation generally did not return solutions for $\delta_l$. Therefore, the 
transition-rotation hybrid of the NPM model is not a viable interpretation of the observation. 


\section{Model Comparison with Observations of PSR B2016+28}
\label{sec:2016}

The solid black lines in the bottom panels of Figures~\ref{fig:models} and~\ref{fig:npmmodel}
show the observed $p$-$\chi$ track covering most of the pulse of PSR B2016+28. Overall, the 
track resembles a canted and elongated figure eight. The track begins at ppb 85, as denoted
by the black dot near the top of the figure eight, where $\chi=0.8^\circ$ and $p=0.146$, and 
ends with similar values of $\chi$ and $p$ at ppb 178. The track initially descends to the 
left, crosses the center of the figure eight, and then proceeds down to its base, where 
$\chi=-41.1^\circ$ and $p=0.087$ at ppb 130. The track then ascends, passing through its 
minimum polarization fraction of $p_m=0.076$ at ppb 137, where $\chi=-32.8^\circ$. Thereafter, 
the track proceeds up and toward the right, again crossing the center of the figure eight. 
As with PSR B1133+16, the $p$-$\chi$ track in PSR B2016+28 is inconsistent with an ideal 
vector rotation or an ideal mode transition. 

The analysis in Section~\ref{sec:1133} showed that interpreting the $p$-$\chi$ variations 
in PSR B1133+16 as a transition-rotation hybrid was not viable, regardless of the 
polarization model used in the interpretation. The same was found to be true for PSR 
B2016+28. Consequently, the following discussion focuses on the mode transition and vector 
rotation cases for each polarization model and excludes the hybrid cases.


\subsection{PCOH model}


\subsubsection{Mode Transition}

\begin{figure}
\plotone{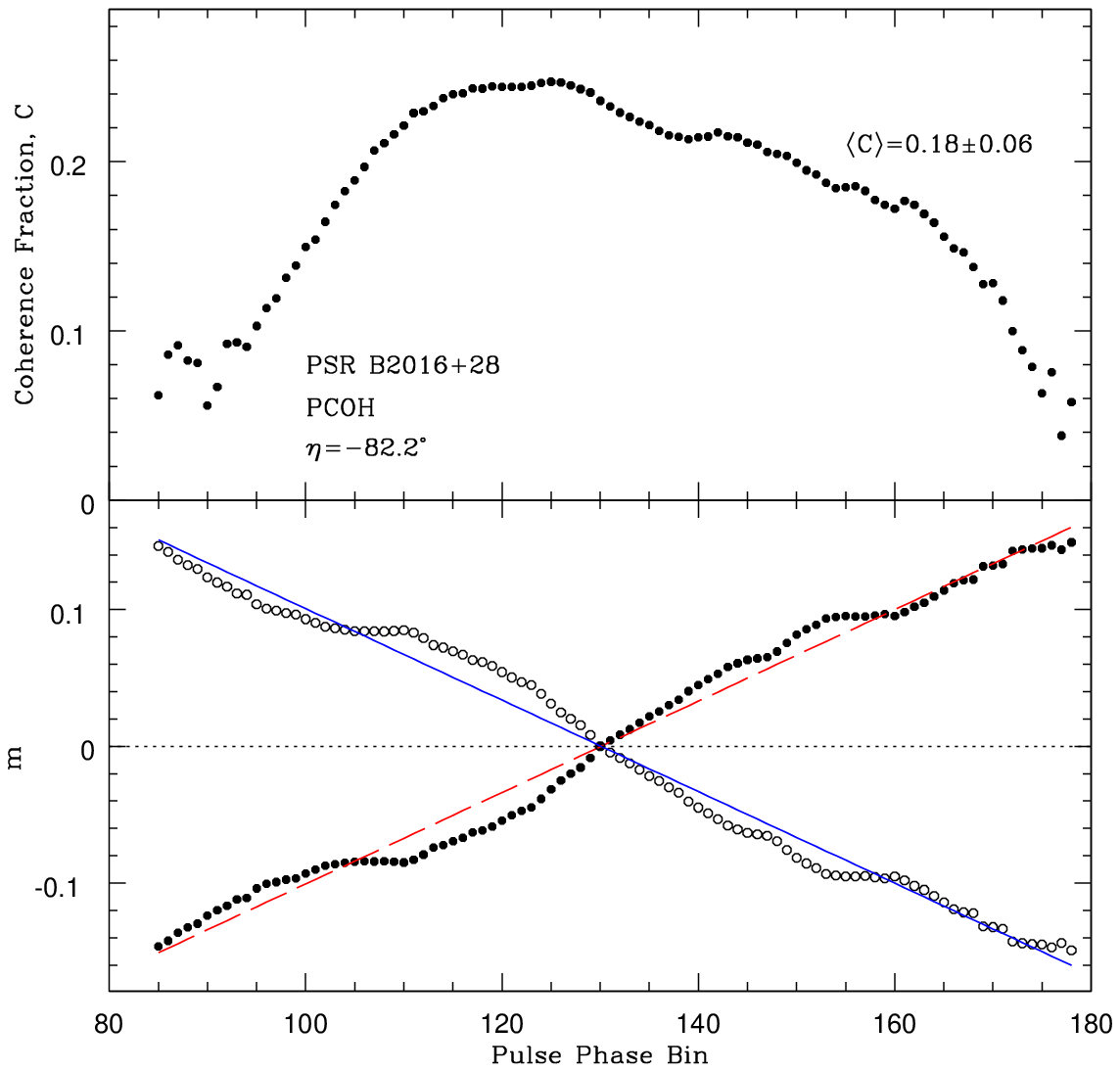}
\caption{Results of the PCOH model applied to PSR B2016+28 assuming the observed variations 
in $p$ and $\chi$ are caused by a mode transition with a varying coherence fraction. The 
calculated coherence fraction is shown in the top panel. The circles in the bottom panel 
are the calculated values of $m$. The lines represent the best fit of the data points to 
a straight line. The mode phase offset used in the calculation was $\eta=-82.2^\circ$.}
\label{fig:mpcoh}
\end{figure}

Figure~\ref{fig:mpcoh} shows the results of interpreting the observations of PSR B2016+28 
as a mode transition in the PCOH model. The fixed value of $\eta$ used in the calculation 
was twice the largest value of the measured EA, $\eta=2\chi=-82.2^\circ$. The top panel of 
the figure shows the calculated coherence fraction is low toward the pulse edges and peaks 
near the pulse center at ppb 125. The mean value of $C$ is $\langle C\rangle = 0.18\pm 0.06$. 
The low values of $C$ are expected given that the polarization fraction observed across the 
pulse is low. The filled and open circles in the bottom panel of the figure show that the 
parameter $m$ varies linearly with ppb. The two lines in the panel represent best fits of 
the data points to straight lines. The fit parameters are listed in Table 2. Both lines 
cross $m=0$ at ppb 130. 

\subsubsection{Vector Rotation}

\begin{figure}
\plotone{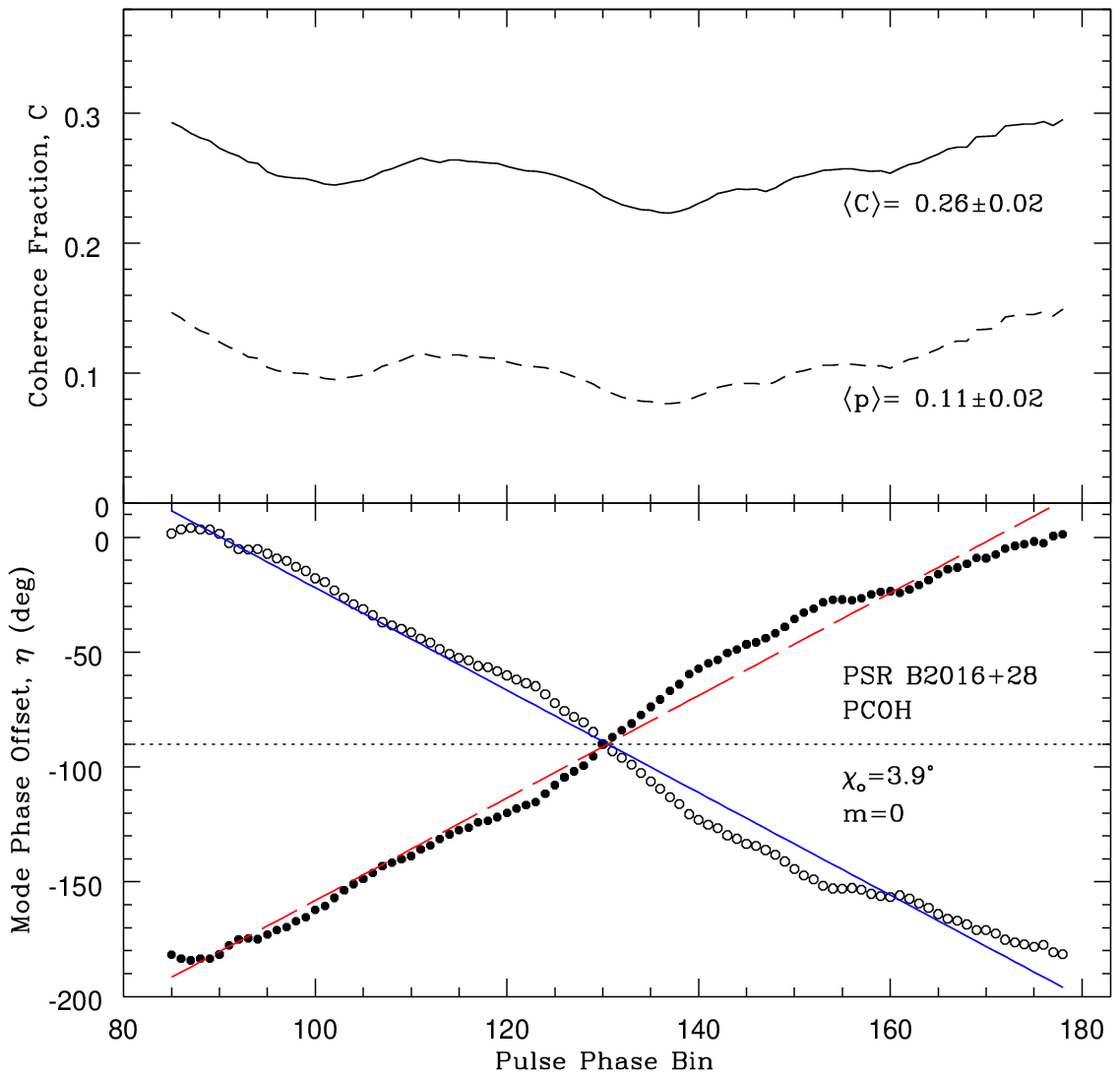}
\caption{Results of the PCOH model applied to PSR B2016+28 assuming the observed variations 
in $p$ and $\chi$ are caused by a vector rotation with a varying coherence fraction. The top 
panel shows the calculated coherence fraction (solid line) and the observed polarization 
fraction (dashed line). The circles in the bottom panel show the calculated values of $\eta$.
The values of $C$ and $\eta$ were calculated assuming the OPMs are elliptically polarized 
with an intrinsic EA of $\chi_o=3.9^\circ$.}
\label{fig:ell2016}
\end{figure}

Similar to PSR B1133+16, the implementation of a vector rotation in the PCOH model is not 
consistent with the observation of PSR B2016+28, because the implementation requires the plane 
of the SC or GC to be perpendicular to the equatorial plane of the Poincar\'e sphere, whereas 
the observed GC is inclined to the plane. An alternate interpretation is the OPMs are 
elliptically polarized, such that the intrinsic EA of the modes is the EA of the normal vector 
to the GC plane, $\chi_n=3.9^\circ$ (see Figure~\ref{fig:EAPA}). In this scenario, the value 
of the parameter $m$ must be zero for the PA-EA pairs to trace a GC. The coherence fraction 
can then be calculated from a combination of $t=1/p$ and Equation~\ref{eqn:t}, and the 
mode phase offset can be calculated from Equation~\ref{eqn:etaell}. The results of the 
calculation are shown in Figure~\ref{fig:ell2016}. The top panel of the figure shows that 
the calculated coherence fraction varies only slightly about a mean value of 
$\langle C\rangle = 0.26$. The variations in $C$ closely track those of the observed 
polarization fraction. The filled and open circles in the bottom panel of the figure show 
that the mode phase offset varies linearly with ppb. Two solutions are possible, because 
the direction of vector rotation cannot be determined in the analysis (OKJ). The two lines 
drawn in the panel represent the best-fit straight lines to the data points. The slopes of 
the lines are $\pm 2.23^\circ/{\rm ppb}$, their intercepts are $-381^\circ$ and $201^\circ$, 
and their correlation coefficients are $\pm 0.994$. Both lines cross $\eta=-90^\circ$ at ppb 
130. 

\subsection{EPC Model}


\subsubsection{Mode Transition}

Interpreting the observation of PSR B2016+28 as a mode transition within the context of 
the EPC model requires the fixed value of $|\chi_e|$ to be less than or equal to the 
largest value of $|\chi|$ observed in the pulse. This constraint gives $\chi_e=-41.1^\circ$ 
and occurs at ppb 130. The results of the mode transition calculation are not shown. The 
resulting values of $\varepsilon$ approach zero toward the pulse edges and increase toward 
the pulse center with a peak at ppb 125 and a mean value of 
$\langle\varepsilon\rangle = 0.05\pm 0.03$. The calculation results also show that the 
parameter $m$ varies linearly with ppb. A straight-line fit of the calculated values of 
$m$ yields fit parameters that are very similar to those obtained with the mode transition 
interpretation of the PCOH model. The fit parameters are listed in Table 2. 


\subsubsection{Vector Rotation}

Interpreting the observation of PSR B2016+28 as a vector rotation within the context of the 
EPC model requires the fixed value of $m$ to be $|m|\le p_m$ everywhere within the pulse. 
This constraint gives $|m|\le 0.076$ and occurs at ppb 130. The results of the vector rotation
calculation are not shown. The calculated values of the EPC relative intensity are low toward 
the pulse edges and peak near the pulse center at ppb 125. The mean value of $\varepsilon$ 
across the pulse is $\langle\varepsilon\rangle = 0.07\pm 0.03$. The calculated values of the 
EPC EA closely track the observed EAs (see the bottom-right panel of Figure~\ref{fig:profiles}).
Unlike the smooth, continuous variations found for the mode phase offset in the PCOH model, 
the EPC EA descends linearly from $\chi_e\simeq 0$ to $\chi_e=-41.1^\circ$, and then 
reverses slope to ascend roughly linearly to $\chi_e\simeq 0$.

\subsection{NPM Model}


\subsubsection{Mode Transition}

\begin{figure}
\plotone{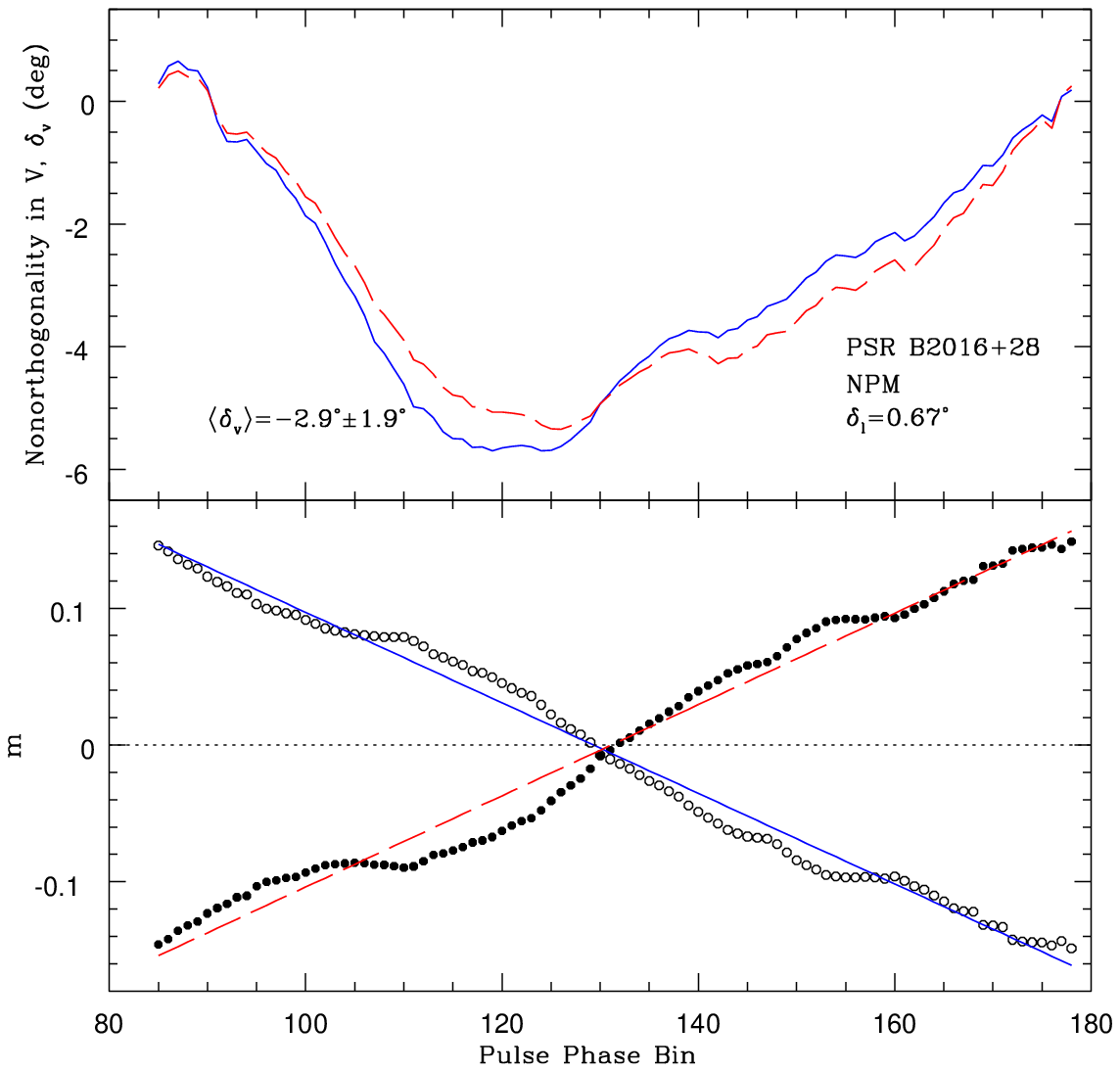}
\caption{Results of the NPM model applied to PSR 2016+28 assuming the observed variations 
in $p$ and $\chi$ are caused by a mode transition with varying nonorthogonality in circular 
polarization, $\delta_v$. The parameter $m$ is shown in the bottom panel, and the value of 
$\delta_v$ is shown in the top panel. The values of $\delta_v$ and $m$ that follow the 
dashed red lines are associated with one another. The same is true of the values of $\delta_v$
and $m$ that follow the solid blue lines. The nonorthogonality in linear polarization used
in the calculation was held constant at $\delta_l=0.67^\circ$.}
\label{fig:npm2016mt}
\end{figure}

Figure~\ref{fig:npm2016mt} shows the results of interpreting the observations of PSR B2016+28 
as a mode transition in the NPM model. The fixed value of $\delta_l$ used in the calculation 
is constrained by $|\sin(\delta_l)|\le p\cos(2\chi)$ everywhere within the pulse. This 
constraint gives $\delta_l=0.67^\circ$ and occurs at ppb 130. The top panel shows that 
the calculated values of $\delta_v$ are low and variable. The mean value of $\delta_v$ 
represented by the blue line is $\langle\delta_v\rangle=-2.9^\circ\pm 1.9^\circ$. As was 
found in PSR B1133+16, the small values of $\delta_l$ and $\delta_v$ returned by the analysis 
are consistent with the model's assumption that the departures from orthogonality are small. 
The filled and open circles in the bottom panel of the figure show that the parameter $m$ 
varies linearly with ppb. The two lines in the panel represent best fits of the data points 
to straight lines. The fit parameters are very similar to those obtained with a mode 
transition in the PCOH and EPC models. The fit parameters are listed in Table 2. 


\subsubsection{Pseudo-vector Rotation}

The bottom panel of Figure~\ref{fig:npm0} shows the results of a calculation that 
assumes the observed variations in $p$ and $\chi$ of PSR B2016+28 are due to a pseudo-vector 
rotation in the NPM model. The parameter $m$ was set to zero in the calculation. The 
calculated values of $\delta_v$ are very similar to what was calculated in the mode
transition case with an identical mean value of 
$\langle\delta_v\rangle=-2.9^\circ\pm 1.8^\circ$. The variations in the calculated values 
of $\delta_l$ track the V-shaped variations in the EA across the pulse. As with PSR 
B1133+16, $\delta_l$ reaches a minimum, and is less than $|\delta_v|$, in the vicinity 
of the EA maximum at ppb 130.

\begin{deluxetable}{ccccc}
\tablenum{2}
\tablecaption{Linear Variation of the Parameter $m$ with Pulse Phase Bin in PSR 
              B2016+28}
\tablehead{\colhead{Model} & \colhead{PCOH} & \colhead{EPC} & \colhead{NPM}}
\startdata
Constant Term & $\eta=-82.2^\circ$ & $\chi_e=-41.1^\circ$ & $\delta_l=0.67^\circ$ \\
Variable Term & $\langle C\rangle = 0.18\pm 0.06$ & $\langle\varepsilon\rangle = 0.05\pm 0.03$
         & $\langle\delta_v\rangle=-2.9^\circ\pm 1.9^\circ$ \\
Slope ($\times 10^3$) & $\pm 3.35$ & $\pm 3.44$ & $ +3.34, -3.31$ \\
Intercept & $\mp 0.435$ & $\mp 0.447$ & $-0.438, +0.428$ \\
Corr. Coeff. & $\pm 0.992$ & $\pm 0.989$ & $ +0.990, -0.994$ \\
\enddata
\end{deluxetable}


\section{Discussion}
\label{sec:discuss}


\subsection{General Comments on the Polarization Models}

OKJ's implementation of the PCOH model assumes the OPMs are partially coherent and linearly 
polarized. The model allows the observations to be interpreted as either vector rotations 
or mode transitions. The changes in mode phase offset that cause the polarization vector 
to rotate can occur as functions of wavelength or pulse longitude due to generalized 
Faraday rotation or Faraday conversion in either the pulsar magnetosphere or wind. If the 
OPMs are incoherent, a mode transition can trace an infinite number of GC geodesics that 
connect the mode polarization vectors on the Poincar\'e sphere, because the vectors are 
antiparallel within the sphere (M24).  However, if the OPMs are partially coherent, the 
transition traces a single, unique geodesic that is defined by the mode vectors and the 
mode phase offset. As shown in Appendix A4 of OKJ, the PCOH model makes allowances for 
both elliptically polarized OPMs and observed departures from mode orthogonality (e.g., 
D. C. Backer \& J. M. Rankin 1980; S84; M03). 

The EPC model is based on suggestions made by S84 and ES04 that the OPMs are accompanied 
by a linearly or circularly polarized emission component (M24). The model adopts the 
more general case of an elliptically polarized emission component. It assumes the OPMs 
are incoherent and linearly polarized. A mode transition in the EPC model traces the 
unique geodesic that is defined by the mode vectors and the EPC EA. The model was 
developed with an emphasis on mode transitions, but can accommodate vector rotations 
by allowing the EPC EA to vary. The current form of the model does not account for 
elliptically polarized OPMs, the partial coherence of the modes, or observed departures 
from mode orthogonality. The physical origin of the EPC and the mechanism responsible 
for altering its EA are not known. This model may not be a realistic representation of 
what is observed given the hypothetical nature of the EPC.

The NPM model assumes the polarization modes are incoherent and nonorthogonal. The 
modes can become nonorthogonal due to mode coupling (D. B. Melrose 1979; S. A. Petrova 
2001) or to large spatial or angular separations of the mode emission beams arising 
from differential refraction (M. C. Allen \& D. B.  Melrose 1982; J. J. Barnard \& 
J. A. Arons 1986; M03). The NPM model explicitly accounts for departures from mode 
orthogonality. It also accounts for elliptically polarized modes in the specific case 
of $\delta_l=0$ (see Equations 12-14 of M24). The model does not account for the partial 
coherence of the modes. Since the mode polarization vectors in the NPM model are not 
orthogonal, a mode transition traces the unique geodesic that connects the vectors on 
the Poincar\'e sphere. Even the slightest departure from mode orthogonality causes the 
transition to trace this unique geodesic. The model was specifically developed for mode 
transitions and consequently is not well suited to interpreting the observations as 
vector rotations. 


\subsection{Distinguishing a Mode Transition from a Vector Rotation}

\subsubsection{Mode Transition}

The distinguishing characteristics of a mode transition and a vector rotation are 
outlined in J. Dyks (2020), DWI, OKJ, and M24. For an ideal mode transition, both 
the polarization fraction and the parameter $m$ vary across the transition. The 
polarization fraction can be as low as $p=0$ if the modes are incoherent, but is 
never equal to zero if the modes are partially or completely coherent. The 
trajectory of the polarization vector created by a mode transition always follows 
a GC on the Poincar\'e sphere and never an SC. Therefore, the resultant vector's 
polarization angles will always follow the GC geodesic that connects the mode 
polarization vectors on the surface of the sphere. The angular extent of a 
complete transition on the sphere is approximately $\zeta=\pi$ (M24). The 
polarization vectors of the OPMs are antiparallel and form a diagonal within the 
sphere. This diagonal resides within the GC plane. If the modes are partially 
coherent, the inclination of the GC plane with respect to the equator of the 
sphere provides a measure of the mode phase offset, $\eta$. If the modes are 
incoherent, the inclination of the plane provides a measure of the modes' 
departure from orthogonality.

\subsubsection{Vector Rotation}

In an ideal vector rotation, the polarization fraction and the parameter $m$ 
remain constant over the rotation. The trajectory of the rotation generally traces 
a portion of an SC, although a GC is possible for a specific geometry 
($\chi_m=\pm\pi/4$). The extent to which the rotation traces a circle in the PCOH 
model, whether partial or complete, depends upon the total change in the mode 
phase offset. The SC axis of rotation is the sphere diagonal formed by the mode 
polarization vectors. Therefore, and unlike the case of a mode transition, the 
diagonal is always perpendicular to the SC plane. Assuming the change in phase 
offset is due to generalized Faraday rotation or Faraday conversion, the 
inclination of the SC plane with respect to the equator of the Poincar\'e sphere 
is an indicator of the polarization of the natural modes of wave propagation 
within the ambient plasma. If the plane is perpendicular to the equator, the OPMs 
are linearly polarized, as in OKJ and DWI, and the particles in the plasma are 
relativistic (V. N. Sazonov 1969; M. Kennett \& Melrose 1998). If the SC plane 
is parallel to the equator, the modes are circularly polarized, and the plasma 
particles are thermal, as with conventional Faraday rotation in the interstellar 
medium. For intermediate inclination angles of the SC plane, the modes are 
elliptically polarized, and the plasma is a mixture of thermal and relativistic 
particles (A. G. Pacholczyk 1973; M. Kennett \& D. Melrose 1998). 


\subsection{Mode Transition or Vector Rotation in PSRs B1133+16 and B2016+28?}

\subsubsection{PSR B1133+16}

The variations in the PA, EA, and polarization fraction across the first PA discontinuity
in PSR B1133+16 are most likely due to a mode transition. The observed PA-EA pairs trace 
approximately one-half of a GC, or a geodesic, as required by a mode transition (top panel 
of Figure~\ref{fig:EAPA}). Additional evidence for the mode transition interpretation 
appears in the pulse profile (left column of Figure~\ref{fig:profiles}), where both the 
linear and total polarization decrease at the locations of the two PA discontinuities. 
Both the linear and total polarization decrease at a mode transition, but in a vector 
rotation, the total polarization remains constant while the linear polarization decreases. 
The interpretation of the observed $p$-$\chi$ variations as a mode transition shows that 
the resulting parameter $m$ varies linearly with ppb, regardless of the polarization model 
used in the interpretation. The slopes, intercepts, and correlation coefficients of the 
straight-line fits to the $m$-ppb data are very similar for all three models (Table 1). 
Consequently, the fits alone do not provide a compelling reason to favor one model over 
the other two. The observed $p$-$\chi$ track is inconsistent with what is predicted by an 
ideal mode transition in the PCOH and EPC models (e.g., see the top panel of 
Figure~\ref{fig:models}). The two models predict a single track, but the observed track 
splits into two branches. As shown in Section~\ref{sec:1133}, departures from the track 
of an ideal mode transition could be attributed to variations in the coherence factor 
of the PCOH model or the EPC intensity of the EPC model. The split $p$-$\chi$ track is 
qualitatively consistent with what is predicted by an ideal mode transition in the NPM 
model. 

Interpreting the $p$-$\chi$ variations as a vector rotation within the context of the 
PCOH and EPC models produces solutions for the mode phase offset and the EPC EA. However, 
$\eta$ and $\chi_e$ do not vary in an organized linear fashion, as the parameter $m$ 
does in the mode transition interpretation. Additionally, the observed $p$-$\chi$ tracks
in the top panels of Figures~\ref{fig:models} and~\ref{fig:npmmodel} do not trace the
single vertical line expected of an ideal vector rotation. Departures from the track of 
an ideal vector rotation could be attributed to variations in the coherence factor or 
the EPC intensity. It is possible for a vector rotation to produce a GC. However, the 
observed GC in the top panel of Figure~\ref{fig:EAPA} is inclined with respect to the 
equatorial plane of the Poincar\'e sphere, which requires the OPMs to be elliptically 
polarized if a vector rotation is to explain the observed $p$-$\chi$ variations.


\subsubsection{PSR B2016+28}

The variations in the PA, EA, and polarization fraction across the pulse of PSR B2016+28
can be interpreted as either a mode transition or a vector rotation. When the observed 
$p$-$\chi$ variations are interpreted as a mode transition, the resulting parameter $m$ 
varies linearly with ppb, regardless of the polarization model used in the interpretation. 
As with PSR B1133+16, the slopes, intercepts, and correlation coefficients of the 
straight-line fits to the $m$-ppb data are very similar for all three models (Table 2). 
The fits alone do not provide a compelling reason to favor one model over the others. The 
PA-EA pairs observed across the pulse resemble a GC geodesic that is expected of a mode 
transition, but one that has been altered by the pulsar's rotation. Support for the mode 
transition interpretation also appears in S84's statistical summary of their single-pulse 
polarization observations of the pulsar. Their Figure 31 shows the individual polarization 
modes occur across the entirety of the pulsar's pulse. One mode dominates on the leading 
edge of the pulse while the other mode dominates on the trailing edge. The frequency of 
occurrence of the modes, as determined from PA histograms, changes systematically across 
the pulse (Figure 1 of M03). The frequency of occurrence is a function of the mode 
intensity ratio and, thus, the parameter $m$ (M. M. McKinnon 2022). Therefore, the 
changing frequency of occurrence is an indication that the parameter $m$ is also varying 
across the pulse. The mode frequency of occurrence would remain constant if the PA-EA
trajectory was caused by an ideal vector rotation, because $m$ is constant in that case.
The separation between the mode PA tracks across the pulse also differs from the 
$\Delta\psi=\pi/2$ expected for OPMs (Figures 1 and 5 of M03), indicating that the 
modes are not orthogonal. The observed $p$-$\chi$ track shown in the bottom panels of 
Figures~\ref{fig:models} and~\ref{fig:npmmodel} is generally inconsistent with what 
is predicted for an ideal mode transition in the polarization models. The minimum 
polarization fraction and the maximum value of the EA do not coincide with one another as 
required by the PCOH and EPC models. In the NPM model, the minimum polarization fraction 
and the maximum EA do not coincide. The $p$-$\chi$ track splits into two branches as 
expected from the NPM model, but the branches close upon themselves to form an elongated 
figure eight. As shown in Section~\ref{sec:2016}, departures from the track of an ideal 
mode transition could be attributed to variations in $C$ for the PCOH model, 
$\varepsilon$ for the EPC model, or $\delta_v$ in the NPM model. 

For a vector rotation to explain the observed PA-EA trajectory in PSR B2016+28, the 
OPMs must be elliptically polarized, because the GC-like feature observed in the 
pulsar is inclined with respect to the equatorial plane of the Poincar\'e sphere 
(see the bottom panel of Figure~\ref{fig:EAPA}). After accounting for the ellipticity 
of the OPMs in the PCOH model, the values of the mode phase offset calculated from the 
observed variations in $p$ and $\chi$ vary linearly with ppb. The calculated values of 
the coherence fraction vary only slightly about a mean value of $\langle C\rangle=0.26$. 
The small variations in $C$ can account for deviations of the observed $p$-$\chi$ track 
from the vertical line expected for an ideal vector rotation. Additional evidence for 
the vector rotation interpretation appears in the pulse profile (right panel of 
Figure~\ref{fig:profiles}). The linear polarization is minimum, and the magnitude of 
the circular polarization is near maximum, where the PA discontinuity and EA excursion 
coincide. The total polarization remains roughly constant over this region of the pulse. 
This behavior is consistent with a vector rotation.


\section{Summary and Conclusions}
\label{sec:conclude}

Existing polarization observations of PSRs B1133+16 and B2016+28 were reanalyzed to
search for GC-like features in their pulse profiles. These features were found in 
both pulsars and accompany the PA discontinuities that are known to occur in their 
profiles. The features were interpreted as a mode transition, a vector rotation, 
and a transition-rotation hybrid within the context of three different polarization 
models. The polarization models are not unique in their ability to represent the 
observations as mode transitions. The hybrid interpretation in each model is 
inconsistent with the observations of both pulsars. The feature observed in PSR 
B1133+16 is most likely a mode transition. The feature observed across the pulse of 
PSR B2016+28 resembles a GC that has been altered by the pulsar's rotation. This GC 
can be interpreted as either a mode transition or a vector rotation. The vector 
rotation interpretation is specific to the PCOH model and requires the OPMs to be 
elliptically polarized. 

PA discontinuities of $\Delta\psi\simeq\pi/2$ often signify mode transitions in 
pulse profiles. Interpreting the discontinuities as transitions between incoherent 
and linearly polarized OPMs can be overly simplistic, because if this interpretation 
was correct, the geodesic that connects the mode polarization vectors would always 
be confined to the equatorial plane of the Poincar\'e sphere. The observations of 
PSRs B1133+16 and B2016+28 show this is not the case, since their GCs are inclined 
with respect to the plane. 

An ideal mode transition causes the PA and EA of a polarization vector to follow
a GC on the Poincar\'e sphere. In general, an ideal vector rotation causes the PA 
and EA to trace an SC on the sphere. The vector geometry of an OPM transition is 
fundamentally different from that of a vector rotation. The antiparallel vectors 
of the OPMs form a diagonal within the Poincar\'e sphere. For an OPM transition, 
the diagonal resides in the GC plane, whereas the diagonal is perpendicular to 
the SC plane for a vector rotation. Each polarization model is characterized by 
three physical parameters. Ideal transitions and rotations occur when two of the 
parameters are fixed, while the remaining parameter varies. When two or all three 
parameters vary, the trajectory of the PA-EA pairs will deviate from the ideal 
GCs and SCs.

The analysis underscores the importance of reporting measurements of the EA. GCs 
or SCs can be revealed when the EA is plotted as a function of the PA or when 
PA-EA pairs are projected on the Poincar\'e sphere. The circles are most likely 
to occur where EA excursions accompany PA discontinuities in pulse profiles.


\appendix

\section{Small and Great Circle Geometry}
\label{sec:circles}

\subsection{Great Circle}

\begin{figure}
\plotone{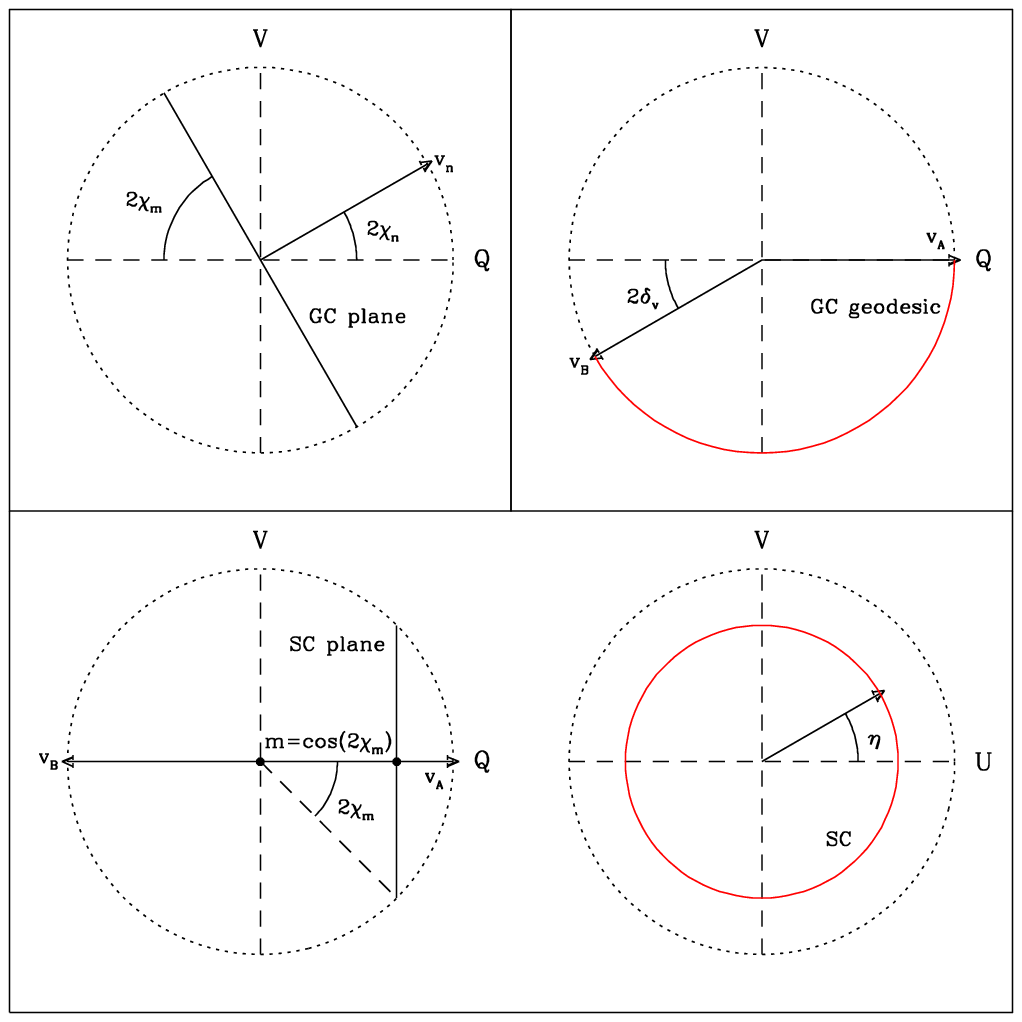}
\caption{Geometry of a GC and an SC arising, respectively, from a mode transition and a 
vector rotation. The top-left panel shows the plane of a GC (solid black line) within the 
Poincar\'e sphere (dotted circle). In this example, the vector that is normal to the GC 
plane, $\rm{v_n}$, has an azimuth of $\phi=2\psi_n=0$ and a latitude of $\lambda=2\chi_n$. 
The top-right panel shows the GC geodesic (red arc) traced by a mode transition from the 
vector of polarization mode A, $\rm{v_A}$, to that of mode B, $\rm{v_B}$, when the 
departures from mode orthogonality are $\delta_l=0$ and $\delta_v$. The bottom panel 
shows the geometry of an SC formed by a vector rotation. The left side of the panel shows 
the unit vectors of the polarization modes reside along the Stokes $Q$ axis of the sphere. 
The intensity of mode A is larger than that of mode B, such that the SC plane (vertical 
black line) is located at $Q=m=\cos(2\chi_m)$. The right side of the panel is a view of 
the sphere along its $Q$ axis and shows the polarization vector rotating about the 
$Q$-axis and tracing an SC (red circle) as the mode phase offset, $\eta$, changes.}
\label{fig:gcsc}
\end{figure}

The geometry of a GC is determined by the intersection of the GC plane with the surface of
the Poincar\'e sphere. The plane is defined by its normal vector, $v_n$. If the latitude 
of the vector is $\lambda=2\chi_n$ and its azimuth is $\phi=2\psi_n$, as illustrated in 
the top-left panel of Figure~\ref{fig:gcsc}, the relationship between the PA and EA of the 
GC is 
\begin{equation}
\tan(2\chi) = \frac{\cos[2(\psi-\psi_n)]}{\tan(2\chi_n)}.
\label{eqn:gc}
\end{equation}
The GC crosses the sphere's equator ($\chi=0$) at PAs of $\psi=\psi_n\pm\pi/4$. The EA maxima
are equal to $|\chi|=\pi/4-|\chi_n|$ and occur at PAs of $\psi=\psi_n$ and $\psi=\psi_n\pm\pi/2$.

A mode transition occurs when the orientation of the observed polarization vector changes 
from the orientation of one polarization mode vector to that of the other mode. The geometry 
of the transition is illustrated in the top-right panel of Figure~\ref{fig:gcsc} for the 
case of the NPM model. The panel shows the vector of one polarization mode, $v_A$, is 
aligned with the Stokes $Q$ axis of the Poincar\'e sphere. The vector of the other mode,
$v_B$, is offset from the $-Q$ axis in latitude by an angle $\lambda=2\delta_v$. The figure
assumes the departure from orthogonality in linear polarization is $\delta_l=0$, such that
both mode vectors reside in the $Q$-$V$ plane of the sphere. The $Q$-$V$ plane is the GC 
plane. The trajectory of the mode transition on the sphere traces the geodesic (red arc in 
the panel) that connects the endpoints of the mode vectors on the sphere's surface. In DWI,
OKJ, and M24, the OPM vectors are parallel and antiparallel to the Stokes $Q$ axis, and 
$\psi_n=\pi/4$.

\subsection{Small Circle}

The SC geometry for a vector rotation in the PCOH model is illustrated in the bottom 
panel of Figure~\ref{fig:gcsc}. As shown on the left side of the panel, the OPMs are 
assumed to be linearly polarized, and their unit polarization vectors, $v_A$ and $v_B$, 
are parallel and antiparallel to the Stokes $Q$ axis of the Poincar\'e sphere. The SC 
plane is perpendicular to the diagonal formed by the mode vectors and is offset from 
the sphere's origin by a distance of $m=\cos(2\chi_m)$\footnote{DWI refer to $\chi_m$ 
as the mixing angle.}. Thus, mode A is the stronger of the two OPMs in this example. 
The vector normal to the SC plane has a PA of $\psi_n=0$ and an EA of $\chi_n=0$. As 
shown on the right side of the panel, a change in the mode phase offset, $\eta$, 
causes the polarization vector to rotate about the mode diagonal and trace a circle 
(red line) in the SC plane. The relationship between the PA and EA for an SC is
\begin{equation}
\cos(2\chi) = \frac{\cos(2\chi_m)}{\cos[2(\psi-\psi_n)]}. 
\label{eqn:sc}
\end{equation}
The SC crosses the sphere equator at PAs of $\psi=\psi_n\pm\chi_m$. Both the maximum 
and minimum values of the EA occur at $\psi=\psi_n$ and are equal to $\chi=\pm\chi_m$. 
The SC becomes a GC when $\chi_m=\pm\pi/4$ (i.e., $m=0$).


\section{Parameter Solutions for Polarization Models}
\label{sec:parms}

The equations for the model parameters written as functions of the observed values of $p$ 
and $\chi$ are listed below. Three cases are considered for each model. Each case holds 
one parameter fixed, while the other two parameters are allowed to vary. The estimated 
value of the fixed parameter can be constrained by the observations, as demonstrated for 
each case.


\subsection{Partially Coherent OPMs}


\subsubsection{Case 1: Mode Transition with Fixed Mode Phase Offset}

For the case of the PCOH model when $\eta$ is fixed while the parameters $m$ and 
$C$ (and thus $t$) vary, Equations~\ref{eqn:EAP} and~\ref{eqn:pp} can be used to 
derive expressions for $m$ and $t$ as functions of the observable quantities 
$p$ and $\chi$:
\begin{equation}
m^2 = \frac{p^2[\sin^2(\eta) - \sin^2(2\chi)]}{\sin^2(\eta)},
\label{eqn:cm}
\end{equation}
\begin{equation}
t^2 = \frac{(1-p^2)\sin^2(\eta) + p^2\sin^2(2\chi)}{p^2\sin^2(2\chi)}.
\label{eqn:ct}
\end{equation}
Since $|m|$ is constrained to lie in the range $0\le |m| \le 1$, the fixed value of 
$|\eta|$ is required to be greater than or equal to $|2\chi|$ everywhere within the 
pulse region of interest. An estimate of the fixed value of $|\eta|$ is then twice 
the maximum value of $|\chi|$ within the pulse region. The values of $m$ and $t$ at 
the ppb with the maximum value of $|\chi|$ are $m=0$ and $t=1/p$.


\subsubsection{Case 2: Vector Rotation with Fixed Mode Intensity Ratio}

For the case when the parameter $m$ is fixed at $m=\cos(2\chi_m)$ (see Figure~\ref{fig:gcsc}), 
the equations for the varying parameters $\sin(\eta)$ and $t$ as functions of the
observable quantities $p$ and $\chi$ are 
\begin{equation}
\sin^2(\eta) = \frac{p^2\sin^2(2\chi)}{p^2-\cos^2(2\chi_m)},
\label{eqn:eta}
\end{equation}
\begin{equation}
t^2 = \frac{\sin^2(2\chi_m)}{p^2-\cos^2(2\chi_m)}.
\label{eqn:tm}
\end{equation}
Since $\sin^2(\eta)$ is constrained to lie in the range of $0\le\sin^2(\eta)\le 1$,
Equation~\ref{eqn:eta} requires $p>\cos(2\chi_m)$ and $p\cos(2\chi)\ge\cos(2\chi_m)$ 
everywhere within the pulse region of interest. The latter inequality places the 
stronger constraint on the estimate of $\chi_m$. Therefore, the value of $\chi_m$, 
and thus $m$, is constrained by the smallest value of the product $p\cos(2\chi)$ 
within the pulse region. In the specific case of $\chi_m=\pm\pi/4$, $m=0$, $t=1/p$, 
and $\eta=2\chi$ everywhere within the region. 

In all three polarization models, a vector rotation occurs when the parameter $m$
is fixed while the other model parameters are allowed to vary. In this instance,
and as indicated by the expressions for the polarization fraction given by 
Equations~\ref{eqn:pp},~\ref{eqn:pe}, and~\ref{eqn:pn}, the fixed value of $|m|$ 
can never exceed the minimum value of $p$ within the pulse region of interest. 
Therefore, the minimum value of $p$ places an upper limit on the value of $m$ 
that can be used in a vector rotation interpretation of the observations. The 
value of $m$ can be more tightly constrained in the case of the PCOH model, as 
discussed in the preceding paragraph.


\subsubsection{Case 3: Transition-rotation Hybrid with Fixed Coherence Fraction}

For the case when $C$ (or $t$) is fixed, the equations for the varying parameters $m$ 
and $\sin(\eta)$ as functions of $p$ and $\chi$ are 
\begin{equation}
m^2 = \frac{p^2t^2-1}{t^2-1},
\label{eqn:hm}
\end{equation}
\begin{equation}
\sin^2(\eta) = p^2\sin^2(2\chi)\left(\frac{t^2-1}{1-p^2}\right).
\label{eqn:heta}
\end{equation}
The two equations place different constraints on the constant value of $t$ and, thus,
$C$. From the allowed values of $|m|$, $t$ must be greater than $1/p$ everywhere within 
the pulse region of interest. Therefore, the first constraint on the fixed value of $t$ 
is a lower limit determined by the minimum value of $p=p_{m}$ within the pulse region,
$t\ge 1/p_m$. At the ppb where $p=p_{m}$, $m=0$ and $\eta=2\chi$. The second constraint 
on $t$ arises from the allowed values of the mode phase offset, $\sin^2(\eta)\le 1$. 
This constraint places an upper limit on the value of $t$ given by
\begin{equation}
t^2\le\frac{1-p^2\cos^2(2\chi)}{p^2\sin^2(2\chi)}.
\label{eqn:tmax}
\end{equation}
At the ppb where $t$ is determined by Equation~\ref{eqn:tmax}, $m=p\cos(2\chi)$ and 
$\eta=\pi/2$.


\subsection{OPMs with an EPC}


\subsubsection{Case 1: Mode Transition with Fixed EPC EA}

For the case of the EPC model when $\chi_e$ is fixed while the parameters $m$ and 
$\varepsilon$ vary, Equations~\ref{eqn:EAE} and~\ref{eqn:pe} can be used to derive 
expressions for $m$ and $\varepsilon$ as functions of the observable quantities 
$p$ and $\chi$:
\begin{equation}
m^2 = \frac{p^2[\sin^2(2\chi_e)-\sin^2(2\chi)]}{[\sin(2\chi_e)-p\sin(2\chi)]^2},
\label{eqn:me}
\end{equation}
\begin{equation}
\varepsilon = \frac{p\sin(2\chi)}{\sin(2\chi_e)}.
\label{eqn:ve}
\end{equation}
From the constraint on the allowed values of $|m|$, Equation~\ref{eqn:me} requires 
$\sin(2\chi_e)>p\sin(2\chi)$ and $|\chi_e|>|\chi|$ everywhere within the pulse
region of interest. The latter inequality is generally the stronger constraint 
of the two, such that the best estimate of $|\chi_e|$ is set by the largest 
value of $|\chi|$ within the pulse region. The values of $m$ and $\varepsilon$ 
at the pulse longitude with the largest value of $|\chi|$ are $m=0$ and 
$\varepsilon=p$.


\subsubsection{Case 2: Vector Rotation with Fixed Mode Intensity Ratio}

When the parameter $m$ is fixed, the equation for the varying parameter $\varepsilon$ 
can be written as a function of $p$ from Equation~\ref{eqn:pe}:
\begin{equation}
\varepsilon=\frac{m^2\pm[p^2-m^2(1-p^2)]^{1/2}}{1+m^2}
\label{eqn:Iep}
\end{equation}
The fixed value of $m$ is constrained by the minimum value of $p$ ($|m|\le p_m$)
within the pulse region of interest. Once $\varepsilon$ is known, $\sin(2\chi_e)$ 
can be calculated from a combination of Equations~\ref{eqn:EAE} and~\ref{eqn:pe}:
\begin{equation}
\sin(2\chi_e) = p\sin(2\chi)/\varepsilon.
\label{eqn:EAp}
\end{equation}
Simple solutions for $\varepsilon$ and $\chi_e$ exist when $m$ is fixed at zero,
specifically $\varepsilon=p$ and $\chi_e=\chi$.


\subsubsection{Case 3: Transition-rotation Hybrid with Fixed EPC Relative Intensity}

When $\varepsilon$ is fixed, the equation for the varying parameter $m$ can be written 
as a function of $p$ from Equation~\ref{eqn:pe}:
\begin{equation}
m^2 = \frac{p^2-\varepsilon^2}{(1-\varepsilon)^2}.
\label{eqn:mepc}
\end{equation}
The value of $\chi_e$ can be calculated from the observed values of $\chi$ and $p$
using Equation~\ref{eqn:EAp}.

As with the hybrid transition-rotation in the PCOH model, the equations for $m$ and 
$\chi_e$ place different constraints on the constant value of $\varepsilon$. From the 
allowed values of $|m|$, $\varepsilon$ must be less than $p$ everywhere within the 
pulse region of interest. The first constraint on the fixed value of $\varepsilon$ is 
then an upper limit determined by $p_{m}$ within the pulse region, $\varepsilon\le p_m$. 
At the ppb where $p=p_{m}=\varepsilon$, note that $m=0$ and $\chi=\chi_e$. The second 
constraint on $\varepsilon$ arises from the allowed values of $\chi_e$. This constraint 
places a lower limit on the value of $\varepsilon$ of $\varepsilon\ge p|\sin(2\chi)|$. 
At the ppb where this constraint is determined, $\chi_e=\pm\pi/4$ and $m$ is given by
\begin{equation}
m=\frac{p\cos(2\chi)}{1-p|\sin(2\chi)|}.
\end{equation}.


\subsection{Nonorthogonal Polarization Modes}


\subsubsection{Case 1: Mode Transition with Fixed Nonorthogonality in Linear Polarization}

For the case of the NPM model when $\delta_l$ is fixed while the parameters $m$ and $\delta_v$ 
vary, Equations~\ref{eqn:EAgen} and~\ref{eqn:pn} can be used to derive equations for $m$ 
and $\delta_v$ as functions of the observable quantities $p$ and $\chi$. The parameter
$m$ is given by
\begin{equation}
m^2 = \frac{2p^2-[1-\cos(2\delta_v)\cos(2\delta_l)]}{1+\cos(2\delta_v)\cos(2\delta_l)}.
\label{eqn:mn}
\end{equation}
The equation for $\delta_v$ is a fourth-order polynomial in $\tan(\delta_v)$. An approximate 
solution for $\tan(\delta_v)$ can be derived by assuming that terms of fifth- and sixth-order 
in products of $\tan(\delta_v)$ and $\sin(\delta_l)$ are negligible:
\begin{equation}
\tan(\delta_v) \simeq \frac{p\sin(2\chi)\cos(\delta_l)
        \{\cos(\delta_l)\pm[p^2\cos^2(2\chi)-\sin^2(\delta_l)]^{1/2}\}}{1-p^2\cos^2(2\chi)}.
\label{eqn:deltav}
\end{equation}
Equation~\ref{eqn:deltav} constrains the fixed value of $\delta_l$ by 
$p\cos(2\chi)>|\sin(\delta_l)|$ everywhere within the pulse region of interest. The 
value of $\delta_l$ is consequently constrained by the location where the product 
$p\cos(2\chi)$ is minimum. Once both $\delta_l$ and $\delta_v$ are known, the parameter 
$m$ can be calculated from Equation~\ref{eqn:mn}. Note that Equation~\ref{eqn:deltav} 
generally produces two different values of $\delta_v$ at each pulse longitude. 

An exact solution for $\tan(\delta_v)$ and $m$ can be derived for the specific case of 
$\delta_l=0$. In this instance, the measured EA and polarization fraction are
\begin{equation}
\tan(2\chi) = \frac{\tan(\delta_v)(1-m)}{\lvert m + \tan^2(\delta_v)\rvert},
\label{eqn:chi0}
\end{equation}
\begin{equation}
p^2 = \sin^2(\delta_v) + m^2\cos^2(\delta_v).
\label{eqn:p0}
\end{equation}
The values of $m$ and $\delta_v$ calculated from Equations~\ref{eqn:chi0} and~\ref{eqn:p0} 
are solutions to quadratic equations. One value of $m$ is always less than or equal to zero 
and is given by
\begin{equation}
m_- = -\frac{p[\cos(2\chi)+p]}{1+p\cos(2\chi)}.
\end{equation}
The other value of $m$ is generally greater than or equal to zero and is given by
\begin{equation}
m_+ = \frac{p[\cos(2\chi)-p]}{1-p\cos(2\chi)}.
\end{equation}
The values of $m_-$ and $m_+$ are equal to one another when $p=0$ or $\chi=\pm\pi/4$.
The values of $\delta_v$ associated with $m_-$ and $m_+$ are, respectively,
\begin{equation}
\tan(\delta_-) = \frac{p\sin(2\chi)}{1+p\cos(2\chi)},
\end{equation}
\begin{equation}
\tan(\delta_+) = \frac{p\sin(2\chi)}{1-p\cos(2\chi)}.
\end{equation}


\subsubsection{Case 2: Pseudo-vector Rotation with Fixed Mode Intensity Ratio}

When the parameter $m$ is fixed, the observed variations in $\chi$ and $p$ are due to 
variations in $\delta_l$ and $\delta_v$. The equations for $\delta_l$ and $\delta_v$ 
as functions of $\chi$ and $p$ are
\begin{equation}
\cos(2\delta_l) = \frac{1}{(1+m)}\frac{1+m^2-2p^2}{[(1-m)^2-4p^2\sin^2(2\chi)]^{1/2}},
\label{eqn:lrot}
\end{equation}
\begin{equation}
\sin(2\delta_v) = \frac{2p\sin(2\chi)}{1-m}.
\label{eqn:vrot}
\end{equation}
The parameter $m$ is constrained by $|m|<p_m$ everywhere within the pulse region of 
interest.


\subsubsection{Case 3: Pseudo-transition-Rotation Hybrid with Fixed Nonorthogonality in 
Circular Polarization}

When $\delta_v$ is fixed while $m$ and $\delta_l$ vary, Equations~\ref{eqn:EAgen} 
and~\ref{eqn:pn} can be used to derive the equation for $\cos(2\delta_l)$ as a function 
of the observable quantities $p$ and $\chi$:
\begin{equation}
\cos(2\delta_l) = \frac{1}{\cos(2\delta_v)}\Biggl\{\frac{\sin^2(2\delta_v)(1-p^2)}
                          {2p\sin(2\chi)[\sin(2\delta_v) - p\sin(2\chi)]} - 1\Biggr\}.
\label{eqn:deltal}
\end{equation}
From the requirement that $\cos(2\delta_l)\le 1$, Equation~\ref{eqn:deltal} places the
following constraint on the fixed value of $\tan(\delta_v)$:
\begin{equation}
\tan(\delta_v)\le\frac{p\sin(2\chi)}{1\pm p\cos(2\chi)}.
\end{equation}
The constraint generally occurs where $p$ is small and the magnitude of $\chi$ is large. 
Therefore, it can be simplified to $\tan(\delta_v)\le p\sin(2\chi)$ everywhere within the
pulse region of interest. Once both $\delta_l$ and $\delta_v$ have been determined, the 
parameter $m$ can be calculated from Equation~\ref{eqn:mn}.



\begin{acknowledgments}

I thank Dan Stinebring for providing the data used in the analysis. The National Radio 
Astronomy Observatory and Green Bank Observatory are facilities of the U.S. National 
Science Foundation operated under cooperative agreement by Associated Universities, Inc.

\end{acknowledgments}


\end{document}